\documentclass[referee,usenatbib]{biom}
\usepackage{amsmath}

\usepackage{amssymb}
\usepackage{float}
\usepackage{mathtools}
\usepackage{graphicx}
\usepackage{caption}
\usepackage{subcaption}
\usepackage{multirow}
\usepackage{colortbl}
\usepackage{comment}
\usepackage{adjustbox}
\usepackage{pdflscape}
\usepackage{rotating}
\floatstyle{plaintop}
\usepackage{xcolor}
\usepackage{fancyhdr}
\usepackage{url}
\usepackage{lineno}
\usepackage{natbib}
\usepackage{enumerate}
\usepackage[shortlabels]{enumitem}
\usepackage{titlesec}
\titlespacing*{\section}{0pt}{0.5\baselineskip}{0\baselineskip}
\titlespacing*{\subsection}{0pt}{0\baselineskip}{0\baselineskip}

\def\bSig\mathbf{\Sigma}

\title[SynDI: Synthetic Data Integration]{A synthetic data integration framework to leverage external summary-level information from heterogeneous populations}

\author{Tian Gu$^{*}$\email{gutian@umich.edu}, 
Jeremy M.G. Taylor, and Bhramar Mukherjee\\
Department of Biostatistics, University of Michigan, Ann Arbor, U.S.A}

\begin{document}


\pagerange{\pageref{firstpage}--\pageref{lastpage}} 
\volume{64}
\pubyear{2022}
\artmonth{May}
\doi{???}

\label{firstpage}

\begin{abstract}
There is a growing need for flexible general frameworks that integrate individual-level data with external summary information for improved statistical inference. External information relevant for a risk prediction model may come in multiple forms, through regression coefficient estimates or predicted values of the outcome variable. Different external models may use different sets of predictors and the algorithm they used to predict the outcome Y given these predictors may or may not be known. The underlying populations corresponding to each external model may be different 
from each other and from the internal study population. Motivated by a prostate cancer risk prediction problem where novel biomarkers are measured only in the internal study, this paper proposes an imputation-based methodology where the goal is to fit a target regression model with all available predictors in the internal study while utilizing summary information from external models that may have used only a subset of the predictors. The method allows for heterogeneity of covariate effects across the external populations. The proposed approach generates synthetic outcome data in each external population, uses stacked multiple imputation technique to create a long dataset with complete covariate information. The final analysis of the stacked imputed data is conducted by weighted regression. This flexible and unified approach can improve statistical efficiency of the estimated coefficients in the internal study, improve predictions by utilizing even partial information available from models that use a subset of the full set of covariates used in the internal study, and provide statistical inference for the external population with potentially different covariate effects from the internal population. 
\end{abstract}




\maketitle


%

\section{Introduction}
\label{s:intro}

Increasingly, researchers are considering incorporating external information from large-scale studies to improve statistical inference rather than using the limited-sized individual-level data that are available to each investigator. It is often easier to have access to the summary information rather than the individual-level data due to restrictions on data privacy and sharing. Therefore, more and more emphasis has been devoted to developing general frameworks that integrate the individual-level data and the summary-level external information in a principled manner. Recent studies on this topic include incorporating external auxiliary information from large data, such as census or population-based biobank data, to improve the statistical inference of the internal study, assuming the model that relates the predictors is fully or partially shared between data sources, also known as transportability \citep*{Bareinboim2013}. Several studies have expanded the initial problem of incorporating summary-level information from a single external data source \citep{Qin2000,Chatterjee2016,Han2019,Estes2017,Gu2019} to multiple external data sources. \citet*{Chatterjee2019} built upon the work of \citet{Chatterjee2016} to a meta-analysis setting, while \citet*{Gu2021} extended the work of \cite{Estes2017} by adaptively weighting multiple empirical Bayes estimators. 

One challenge is accommodating the heterogeneity among different data sources, ignoring which would lead to potential estimation bias and misleading inference during data integration. Efforts have been made to address this issue. The framework proposed by \citet{Gu2021} assigns larger weights to the more compatible external data sources to incorporate valid supplementary information into the internal study. \citet{Chen2020} used a penalty function to identify the difference of aggregate information among data sources. \cite{Yang2020} employed a sensitivity parameter to quantify such systematic differences. Moreover, there could be other sources of information variation across the models. For example, different external studies may use different subsets of covariates, the underlying prediction model may be parametric or constructed by machine learning approaches, and the summary-level information may consist of estimated regression coefficients or of fitted predictions.

Although some of the existing approaches have considered the heterogeneity across data sources, the main focus has been on improving the statistical efficiency of the internal/main dataset with little attempt to make statistical inference on the external populations or allowing heterogeneous covariate effects across data sources. \citet*{Fei2012} proposed a joint estimating procedure to merge longitudinal datasets while allowing different study-specific coefficients. The meta-analysis approach proposed by \citet*{Chatterjee2019} used a generalized method of moments to estimate study-specific effects but only allows covariates that were measured in at least one of the external studies.

In this study, we consider the situation where moderately sized individual data is available from the internal study, and there are K populations (K $\rm \ge 1$), each of which provides some information about the relationship between the same outcome and a slightly different set of predictors. The goal is to fit a target regression model with all available predictors in the internal study while utilizing summary information from external models that may have used only a subset of the predictors. We propose a novel data integration and analysis strategy combining and extending two existing methods, the synthetic data method \citep{Gu2019} and the stacked imputation method \citep{beesley2021stacked}. Our innovation includes first extending \citet{Gu2019} to convert multiple external models into synthetic data, and then proposing a stacked imputation strategy that includes \citet{beesley2021stacked} as a special case, and which further allows for heterogeneous covariate effects across populations. This flexible and unified data integration approach attains the following four objectives: (i) incorporating supplementary information from a broad class of externally fitted predictive models or established risk calculators based on parametric regression or machine learning methods, as long as the external model can generate predicted outcomes given covariates; (ii) improving statistical efficiency of the estimated coefficients in the internal study; (iii) improving predictions by utilizing even partial information available from prediction models that uses a subset of the full set of covariates used in the internal study; (iv) making statistical inference for the external population with potentially different covariate effects from the internal population; and (v) we develop open-source software to help users to easily implement the proposed method through an R package. 

The rest of the paper is organized as follows: In Section \ref{motivation}, we introduce the motivating example where two new biomarkers that are measured in the internal study may be able to improve the risk prediction of high-grade prostate cancer. In Section \ref{method}, we introduce the proposed methodology and develop an user-friendly software through an R package to implement it. In Section \ref{realdata}, we apply the proposed strategy to the motivating example, where we build an expanded risk model to predict high-grade prostate cancer borrowing information from two existing risk prediction models.  In Section \ref{simulation}, we evaluate the performance of our proposed approach in a simulation study.  Concluding remarks are presented in Section \ref{Discussion}.

\section{High-grade Prostate Cancer Risk Prediction} \label{motivation}
Gleason grade is a score that is used to measure the severity of the prostate cancer at the time of diagnosis, where higher level represents worse prognosis. Several well-established risk calculators are available to predict the risk of high-grade prostate cancer (Gleason score over 6) in men undergoing a biopsy. For example, the Prostate Cancer Prevention Trial risk calculator (PCPThg) established from a United States population \citep{Thompson2006} and the European Randomized Study of Screening for Prostate Cancer risk calculator 3 (ERSPC) established from a European population \citep{ERSPC2012} each used slightly different predictors to predict the same outcome through logistic regression models. 

Specifically, PCPThg and ERSPC both used prostate-specific antigen level (PSA) and digital rectal examination findings (DRE) as predictors, and PCPThg also used age, race (African American or not) and prior biopsy results while ERSPC additionally used transrectal ultrasound prostate volume (TRUS-PV) with fitted models given by:
\begin{itemize}
    \item PCPThg: $\rm logit(p_i)=-3.69+0.89log_2(PSA_i)+DRE_i+0.03Age_i+0.96Race_i-0.36Biopsy_i$;
    \item ERSPC: $\rm logit(p_i)=-3.16+1.18log_2(PSA_i)+1.81DRE_i-1.51log_2(\text{TRUS-PV}_i)$,
\end{itemize}
where $\rm p_i$ is the probability of observing high-grade prostate cancer for subject $i$. Both of these risk calculators were established from large datasets (18882 men in PCPThg and 3624 men in ERSPC) and have been evaluated in independent validation datasets.

In addition, new molecular biomarkers have been identified to be predictive of prostate cancer, two examples of which are prostate cancer antigen 3 (PCA3) and TMPRSS2:ERG (T2:ERG) gene fusions \citep{Tomlins2015, Truong2013}. In an internal dataset provided by \citet{Tomlins2015}, we have 678 male patients who have complete data of all the predictors used in PCPThg and ERSPC, as well as PCA3 and T2:ERG gene fusions. Therefore, it is natural to consider using both the new biomarkers and the existing prostate cancer predictors to build an expanded risk calculator aiming for improved prediction performance. Since the data containing such new biomarkers are usually available on moderately sized samples with limited predictive power if being analyzed alone, incorporating established model information, such as the information from PCPThg and ERSPC, can potentially improve the prediction accuracy and is of great interest for precision prevention for men at risk of high grade prostate cancer. Prior studies have shown promising risk discrimination by combining PCA3 and/or T2:ERG with PCPThg \citep{Tomlins2015, Cheng2019, Gu2021}.

In addition to the internal dataset, there is an independent dataset of size 1174, which was collected from seven community clinics throughout the United States. This dataset will be used for validation. The baseline characteristics of each dataset is listed in Web Supplemental Section 1. From this we can see considerable differences in the distributions of some of the baseline variables across the datasets. We note especially that the population for the PCPThg study has lower risk, based on lower PSA values, and the prevalence of high grade prostate cancer is also much lower in this study. We would like to propose a more flexible framework which firstly expands the existing prediction model to include new predictors in the internal population, and secondly boosts the estimation efficiency of the coefficients by incorporating externally fitted models, accounting for population heterogeneity.

\section{Models and Methods} \label{method}
\subsection{Notation} \label{notation}
Let Y denote the outcome of interest, which can be continuous, binary or of other types. Consider $\rm \mathbf{X}$ a set of P routinely measured predictors and $\rm \mathbf{B}$ a set of Q new predictors, e.g., newly discovered biomarkers, where $\rm \mathbf{B}$ is only available in the internal study (i.e., $\rm \mathbf{B}$ are unmeasured predictors in all of the external studies). Let $\rm \mathbf{I}= (I_1,...,I_K)$ denotes the indicator vector representing $\rm K$ external populations and $\rm I_0$ represents the internal study. 

We discuss the problem under the assumption that $\rm Y|\mathbf{X,B}$ is linear in $\rm \mathbf{X}$ and $\rm \mathbf{B}$. We consider the case where a moderate dataset of size n with complete predictors $\rm Y, \mathbf{X}$ and $\rm \mathbf{B}$ is available to us from the internal study. For each external population k $\ge$ 1, a well-established reduced model for the same outcome $\rm Y$ is also available, each of which may use a slightly different set of predictors $\rm \mathbf{X}_k$, a subset of $\mathbf{X}$. For example, if linear regression is used, the prediction model may look like: $\rm E(Y|\mathbf{X}_k)
        = \boldsymbol{\rm X}_k \boldsymbol{\rm \beta}_k = \rm \beta_0 + \sum_{p=1}^{P_k}\beta_p X_p$,
where $\rm P_k \le P$ is the dimension of $\rm \mathbf{X}_k$. We do not have access to the underlying individual-level data that was used to fit the external model but only have the summary information. This summary information can come in different forms that we summarize into two categories:
\begin{enumerate}[label={\bfseries Category \arabic*:},wide = 0pt, leftmargin = 3em]
    \item Directly available in the form of an externally fitted parametric regression model, along with the estimated model parameters $\rm \hat{\boldsymbol{\beta}}_k$;
    \item Any parametric or non-parametric model without knowing the exact form, e.g., an established risk calculator that can provide the risk probability $\rm P(Y=1|\mathbf{X}_k)$ given any $\rm \mathbf{X}_k$.
\end{enumerate}
The target model of interest is a generalized linear model (GLM) including all $\rm \mathbf{X}$ and $\rm \mathbf{B}$ that allows for population-specific intercepts and X coefficients in the following form:
\begin{equation}\label{general Y|XBS}
        g[\rm E(Y|\mathbf{X, B, I})] = \rm \underbrace{\rm \gamma_0 + \sum_{k=1}^K \gamma_0^{k}I_k}_{\text{population-specific intercepts}} + \sum_{p=1}^P \gamma_{X_p}X_p + \underbrace{\rm \sum_{k=1}^K \sum_{p=1}^P \gamma_{X_p}^{k}X_pI_k}_{\text{heterogeneous X effects}} + \sum_{q=1}^Q \gamma_{B_q} B_q,
    \end{equation}
where $g$ is the known link function. If one does not want to incorporate information from external studies, model (\ref{general Y|XBS}) will reduce to the direct regression using internal data only:
    \begin{equation}\label{general Y|XBS=0}
        g[\rm E(Y|\mathbf{X, B}, I_0)] = \rm \gamma_0 + \sum_{p=1}^P \gamma_{X_p} X_p + \sum_{q=1}^Q \gamma_{B_q} B_q.
    \end{equation}
Our goal is to obtain the estimates of $\boldsymbol{\gamma}$ in model (\ref{general Y|XBS}). We assume that model (\ref{general Y|XBS}) is correctly specified, which indicates that each external population can potentially differ in intercept and $\rm \mathbf{X}$ covariate effect as long as those $\rm \mathbf{X}$'s were used in the external model, while for covariates that are unmeasured in the $\rm k^{th}$ model, we assume the covariate effects are the same as the internal study. In addition, model (\ref{general Y|XBS}) is a saturated model allowing all intercepts and $\rm \mathbf{X}$ covariates to differ across populations. In practice, given prior knowledge, we could reduce the number of unknown parameters to estimate by forcing some $\rm \gamma_0^{k}$ and/or $\rm \gamma_{X_p}^{k}$ to be zero. For example, a special case of model (\ref{general Y|XBS}) is a logistic regression model that only allows population-specific intercepts, representing different prevalences and the same covariate effects in each population:
\begin{equation}\label{target Y|XBS}
    logit[\rm Pr(Y=1|\mathbf{X, B, I})] = \underbrace{\rm \gamma_0+ \sum_{k=1}^K \gamma_0^{k}I_k}_{\text{population-specific intercepts}} + \sum_{p=1}^P \gamma_{X_p} X_p + \sum_{q=1}^Q \gamma_{B_q} B_q.
\end{equation}
We assume that the external models $\rm Y|\mathbf{X}_k$'s are the best-fitted models in the class of the reduced models that was considered, but this class of reduced models may not contain the true distribution of $\rm Y|\mathbf{X}_k$. One such example is when the full model is the logistic model shown in equation \ref{target Y|XBS}, but the true distribution for $\rm Y|\mathbf{X}_k$'s are not logistic models as collapsibility does not hold for the logit link. We consider the fitted logistic model as the best-fitted model in the class. 

\subsection{Proposed Synthetic Data Integration and Analysis Framework (SynDI)} \label{framework}
\begin{sidewaysfigure}
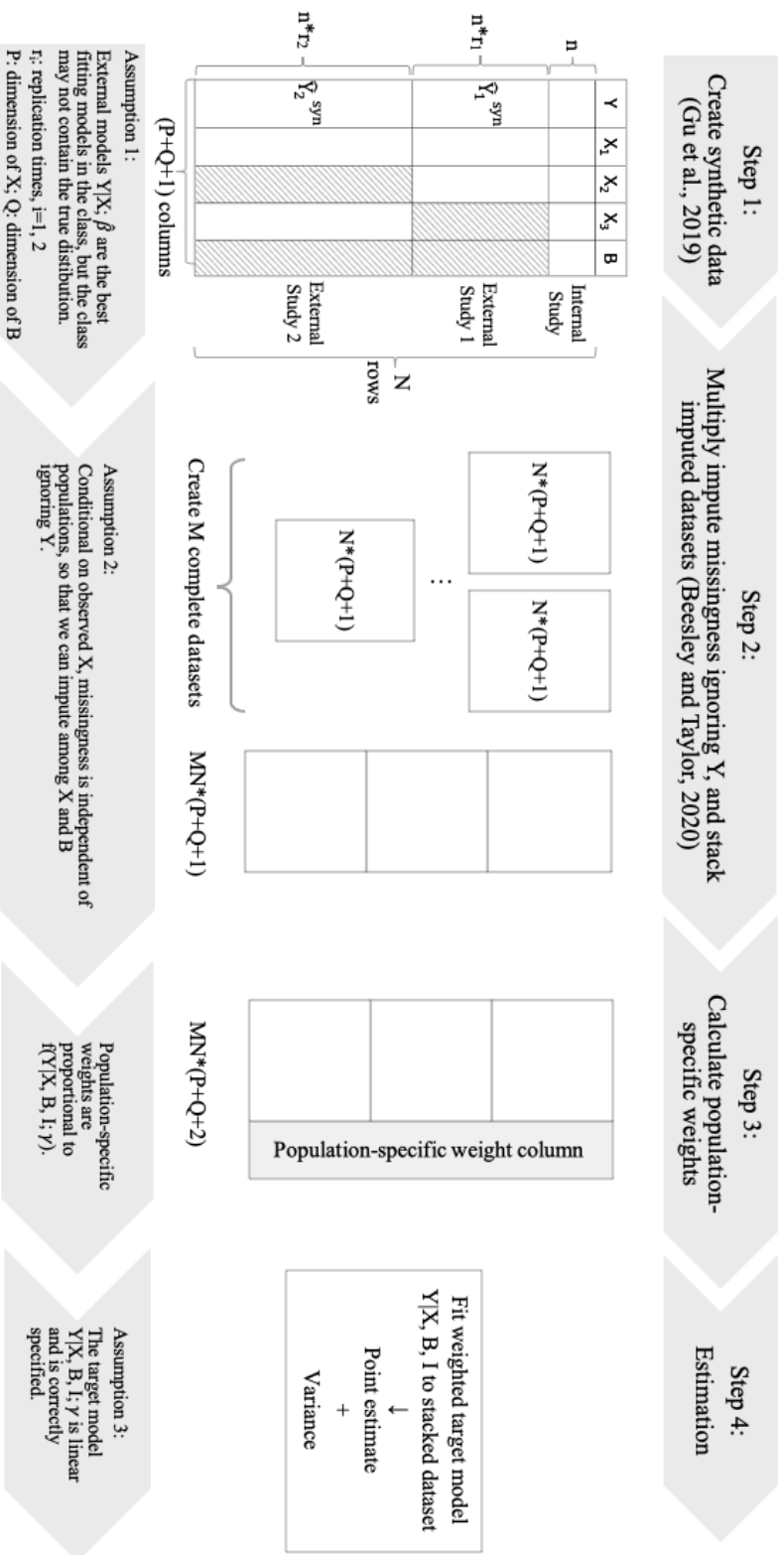

\centering
\adjustimage{width=.9\linewidth, center=1\linewidth}{flowchart}
\caption{Diagram of the proposed strategy.} \label{flowchart}
\end{sidewaysfigure}

We propose a synthetic data integration and analysis strategy, SynDI, that allows for heterogeneous $\rm \mathbf{X}$ covariate effects across the external populations, by first generating synthetic data for each external population, then using a stacked multiple imputation strategy to impute the missingness, and finally analyzing the imputed data through a heterogeneity-weighted regression. Figure \ref{flowchart} illustrates the proposed four-step strategy, along with the required assumptions in each step:

\noindent\underline{\textbf{Step 1: Convert each external summary-level information into a set of synthetic}} \\ \underline{\textbf{data.}} We propose to leverage the external data information through a synthetic data approach, where the idea is to generate one set of synthetic data, ($\rm \hat{Y}_k^{syn}, \mathbf{X}_k^{syn}$), for each of the $\rm k^{th}$ external study such that the distribution of $\rm \hat{Y}_k^{syn}|\mathbf{X}_k^{syn}$ mimics the estimated conditional distribution $\rm Y|\mathbf{X}_k; \hat{\boldsymbol{\beta}}_k$ in the $\rm k^{th}$ study. These synthetic data will be combined with the internal data to jointly estimate the target parameter $\boldsymbol{\gamma}$ in model (\ref{general Y|XBS}). Since we want to borrow information from the conditional distribution $\rm Y|\mathbf{X}_k$ in the external population, the marginal distribution of $\rm X$ in the synthetic data does not need to match the external study. Therefore, we create synthetic covariates following the same distribution as the internal covariates and the synthetic outcome is generated from the external model estimates.

Specifically, to create synthetic data for external study k, we first generate the synthetic covariates $\rm \mathbf{X}_k^{syn}$ by replicating $\rm r_k$ times the internal data, where $\rm r_k$ is some positive integer. We will discuss the choice of $\rm r_k$ in the following paragraph. We then simulate the synthetic outcome $\rm \hat{Y}_k^{syn}$ using $\rm \mathbf{X}_k^{syn}$ and the external summary information $\rm Y|\mathbf{X}_k; \hat{\boldsymbol{\beta}}_k$. Unmeasured predictors in the external populations (all $\rm \mathbf{B}$ and some $\rm \mathbf{X}$'s) will be treated as missing data. Illustrated in Figure \ref{flowchart} step 1, assuming there are $\rm K=2$ external studies, since the external study 1 used $\rm X_1$ and $\rm X_2$ to predict Y, we first replicate the observed ($\rm X_1$, $\rm X_2$) in the internal study $\rm r_1$ times to create $\rm X_1^{syn}$ of size $\rm n \times r_1$; we then utilize the summary information $\rm Y|X_1, X_2; \hat{\boldsymbol{\beta}}_1$ from external model 1 to generate the synthetic outcome values $\rm \hat{Y}_1^{syn}$ given $\rm X_1^{syn}$; and lastly, the unmeasured predictors $\rm X_3$ and $\rm B$ will remain missing. Similarly for the external study 2, we replicate the observed ($\rm X_1$, $\rm X_3$) as $\rm X_2^{syn}$, and create synthetic values $\rm \hat{Y}_2^{syn}$. After combining the synthetic data with the target data, this combined dataset is of size N $\rm \times$ (P+Q+1), where $\rm N=n\times(1+r_1+r_2+...+r_K)$.

Step 1 converts the external models' summary information into synthetic individual-level data, allowing for analyzing the observed internal data together with the synthetic external data. Similar ideas of creating synthetic data have been considered in survey methodologies \citep{reiter2002satisfying, raghunathan2003multiple, reiter2012inferentially}, causal inference \citep{tan2021tree} and high-dimensional transfer learning \citep{Gu2022SynTL}. We follow the same procedure of creating synthetic data as introduced in \citet{Gu2019}, and extend them to accommodate the case of multiple external models. 

\cite{Gu2019} provided theoretical justification in special cases to show that the synthetic data method is equivalent to a constrained semi-parametric maximum likelihood approach \citep{Chatterjee2016}, and it assumed the external models were the best-fitted models in the class, but the class may not contain the true distribution (Assumption 1). They showed in finite sample size, the larger the number of replicates, $\rm r_k$, in each synthetic dataset, the more precision gain in the estimated coefficient of $\rm X$; when $\rm r_k$ goes to infinity, the precision gain by incorporating the $\rm k^{th}$ external model will converge to a constant. In practice, it is reasonable to set the synthetic data size, i.e., $\rm n \times r_k$, similar to the external study's actual study size. We will assess the performance of SynDI by varying $\rm r_k$ in the simulation studies in Section \ref{simulation}.

\noindent\underline{\textbf{Step 2: For the combined dataset created in step 1, multiply impute the missing}}\\\underline{\textbf{covariates without using the outcome Y to create M imputed datasets.}} To handle the missing covariates in the combined dataset created in step 1, we propose to multiply impute the missingness through traditional multiple imputation by chained equation (MICE) while not involving the use of the outcome Y when creating the imputed covariates. The M imputed datasets will then be stacked together, constituting a stacked long dataset of size MN $\rm \times$ (P+Q+1). 

The idea of stacked imputation was proposed in \cite{beesley2021stacked} to flexibly employ multiple imputation when the outcome is hard to adjust for in the imputation model, such as in the situation where the outcome is a censored survival time. The idea is to impute the missingness using covariates only and analyze a weighted regression using weights proportional to probabilities derived from the outcome model, where in the following theoretical justification we briefly explain why we can achieve the same result as if including the outcome in the imputation. Specifically, to fit a regression model $\rm Y|\mathbf{X, B}$ with missingness in $\rm \mathbf{X}=(\mathbf{X}_{mis},\mathbf{X}_{obs})$ and $\rm \mathbf{B}$, the authors decomposed the MICE imputation model distribution into two terms, $\rm f(\mathbf{X}_{mis},\mathbf{B}|Y, \mathbf{X}_{obs}) \propto f(\mathbf{X}_{mis}, \mathbf{B}|\mathbf{X}_{obs})f(Y|\mathbf{X}_{mis},\mathbf{X}_{obs},\mathbf{B})=\underbrace{\rm f(\mathbf{X}_{mis},\mathbf{B}|\mathbf{X}_{obs})}_{\text{term 1}} \underbrace{\rm f(Y|\mathbf{X,B})}_{\text{term 2}}$. They proposed to first use term 1 ignoring Y to multiply impute the missingness, then stack the imputed datasets, and finally use term 2, that is proportional to the model of interest, to calculate a set of weights for a final weighted regression model. The weights are estimated through $\rm f(Y|\mathbf{X,B}; \boldsymbol{\hat{\gamma}}_{I_0})$, where $\rm \boldsymbol{\hat{\gamma}}_{I_0}$ is the direct regression estimates of model (\ref{general Y|XBS=0}).

We apply this idea of stacked imputation in step 2 and extend the weight calculation to allow for heterogeneous $\rm \mathbf{X}$ covariate effects in step 3 below. Since our target model is $\rm Y|\mathbf{X, B, I}$, our imputation model becomes $\rm f(\mathbf{X}_{mis},\mathbf{B}|Y, \mathbf{X}_{obs}, \mathbf{I}) \propto \underbrace{\rm f(\mathbf{X}_{mis},\mathbf{B}|\mathbf{X}_{obs}, \mathbf{I})}_{\text{term 1}} \underbrace{\rm f(Y|\mathbf{X,B,I})}_{\text{term 2}}$. To implement multiple imputation in the block-wise missing structure, where some covariates are completely unobserved in one population, we assume term 1 is the same across populations, i.e., $\rm f(\mathbf{X}_{mis},\mathbf{B}|\mathbf{X}_{obs}, \mathbf{I}) = f(\mathbf{X}_{mis},\mathbf{B}|\mathbf{X}_{obs})$ (Assumption 2). Detailed explanation of Assumption 2 can be found in the Web Supplemental Section 2 and will be discussed later.

\noindent\underline{\textbf{Step 3: Calculate population-specific weights for each subject in the stacked long}}\\ \underline{\textbf{dataset.}} The weights in term 2 above are proportional to the target model distribution $\rm f(Y|\mathbf{X,B,I}; \boldsymbol{\gamma})$, where initial estimates of $\boldsymbol{\gamma}$ is needed to calculate the population-specific weights. Recall the internal estimates is $\rm \boldsymbol{\hat{\gamma}}_{I_0}$, and denote the external population estimates as $\rm \boldsymbol{\hat{\gamma}}_{I_k}$. In this step, we extend the original weights proposed in \cite{beesley2021stacked} to population-specific weights proportional to $\rm f(Y|\mathbf{X,B, I})$ for the internal population $\rm I_0$ and each external population $\rm I_k$, where \cite{beesley2021stacked} reduces to a special homogeneous-population case of ours when $\rm I_0=I_1=...=I_K$.

Specifically, for the internal population, we directly compute weights proportional to $\rm f(Y|\mathbf{X,B}; \boldsymbol{\hat{\gamma}}_{I_0})$. For the $\rm k^{th}$ external population, $\rm \boldsymbol{\hat{\gamma}}_{I_k}$ from model $\rm Y|\mathbf{X,B},I_k; \hat{\boldsymbol{\gamma}}_{I_k}$ is not directly available since we only have the summary information on the reduced model $\rm Y|\mathbf{X}_k; \hat{\boldsymbol{\beta}}_k$. Therefore, we treat the unmeasured covariates, $\rm \mathbf{X}_{(-k)}$ and $\mathbf{B}$, as the omitted covariates, assuming they have the same effects as the internal population. \cite{Neuhaus1993} argued that $\rm \hat{\boldsymbol{\beta}}_k$ would be biased due to omitting the covariates, and derived the algebraic relationship between $\rm \hat{\boldsymbol{\beta}}_k$ and the true effect of $\rm \mathbf{X}_k$. Following \cite{Neuhaus1993}, We use the internal data and $\rm \hat{\boldsymbol{\beta}}_k$ to estimate the true effect of $\rm \mathbf{X}_k$, and thus compute $\rm \hat{\boldsymbol{\gamma}}_{I_k}=(\hat{\gamma}_0^{I_k}, \hat{\boldsymbol{\gamma}}_{X_k}^{I_k}, \hat{\boldsymbol{\gamma}}_{X_{(-k)}}, \hat{\boldsymbol{\gamma}}_B)^T$, where $\rm \hat{\gamma}_0^{I_k}$ and $\rm \hat{\boldsymbol{\gamma}}_X^{I_k}$ are bias-corrected intercept and $\rm \mathbf{X}$ coefficients from $\rm \boldsymbol{\hat{\beta}}_k=(\hat{\beta}_0, \boldsymbol{\rm \hat{\beta}}_{X_k})^T$ while $\rm \hat{\boldsymbol{\gamma}}_{X_{(-k)}}$ and $\rm \hat{\boldsymbol{\gamma}}_B$ are estimated coefficients from $\rm \boldsymbol{\hat{\gamma}}_{I_0}$. More details can be found in Web Supplemental Section 3, along with two examples to derive the initial estimates for linear and logistic regression, respectively.  

As described in Section \ref{notation}, the summary information will be available in the form of either parameter estimates $\rm \boldsymbol{\hat{\beta}}_k$ (Category 1) or a risk calculator of unknown form that has the ability to estimate the probability of Y=1 given $\rm \mathbf{X}_k$ (Category 2). In the case of Category 1, we can directly use $\rm \boldsymbol{\hat{\beta}}_k$. In the case of Category 2 where $\rm \boldsymbol{\hat{\beta}}_k$ does not exist, we follow the same procedure as in Category 1 but use an approximation $\rm \boldsymbol{\hat{\beta}}^{syn}_k$ instead. Specifically, we first create a large size of synthetic data ($\rm \hat{Y}_k^{syn}$, $\rm \mathbf{X}^{syn}_k$) as described in step 1, and then we fit a GLM $\rm \hat{Y}_k^{syn}|\mathbf{X}^{syn}_k$ with main effect using only the synthetic data, from which we obtain $\rm \boldsymbol{\hat{\beta}}^{syn}_k$. Further assessment can be found in simulation II of Section \ref{simulation}.


\noindent\underline{\textbf{Step 4: Estimate the target parameter $\rm \boldsymbol{\gamma}$ of model (\ref{general Y|XBS}) through a weighted GLM}}\\\underline{\textbf{using the stacked long dataset.}} Given the stacked long dataset obtained in step 2 and the weights calculated in step 3, we can jointly fit a weighted regression of the target model (\ref{general Y|XBS}) applied to the stacked long dataset to obtain the final point estimates of $\boldsymbol{\gamma}$. 

To measure the variation of $\rm \boldsymbol{\hat{\gamma}}$, we cannot direct to use the variance estimator from the weighted regression nor the sandwitch estimator, because they are biased as they do not account for the between-imputation variation \citep{wood2008should, beesley2021stacked}. \citet{beesley2021accounting,beesley2021stacked} proposed three variance estimators. However, our numerical assessment reveals that the bias of these existing variance estimators increases with the synthetic data size, which conflicts with the fact that larger synthetic data size will provide more precise point estimates. Therefore, we propose a new bootstrap procedure by resampling the internal data and repeating steps 1-4, which we show to have stable performance and outperforms the existing variance estimators.

\section{Application to the Prostate Cancer Data} \label{realdata}
We apply SynDI to predict the risk of high-grade prostate cancer using an internal dataset containing patients from three United States academic institutions \citep{Tomlins2015} and two external risk calculators, PCPThg \citep{Thompson2006} and ERSPC \citep{ERSPC2012} introduced in Section \ref{motivation}.

A total of 678 patients who had complete data were included in the internal data set provided by \citet{Tomlins2015} (the prevalence of high-grade prostate cancer is 26.4\%). In the internal individual-level data, in addition to all the predictors used in PCPThg and ERSPC, we also have data on two new biomarkers, PCA3 and T2:ERG gene fusions mentioned in Section \ref{motivation}. Therefore, in Table \ref{Table_realdata}, we present the results of two target models using a total of eight predictors, including these two new biomarkers. One model only allows the intercept to be different across populations (``different intercept only" model corresponds to model [\ref{target Y|XBS}]), and another flexible model allows all possible covariates that are used in the external populations to have population-specific effects (``different intercept and covariates" model that corresponds to model [\ref{general Y|XBS}]). The detailed model forms can be found in the table legend.  In Table \ref{Table_realdata}, we present the point estimates (estimated standard deviation, SE) of each predictors. Using the validation dataset of $\rm N_{test}=1174$, we also present the area under the curve (AUC) and the scaled Brier Score (BS) defined as $\rm \sum_{i=1}^{N_{test}}(Y_i-\hat{p}_i)^2 / \sum_{i=1}^{N_{test}}(Y_i-\bar{Y})^2$, where $\rm Y_i$ is the observed outcome for the $i$-th patient in the validation data, and $\rm \bar{Y}=\frac{1}{N_{test}}\sum_{i=1}^{N_{test}}Y_i$.

\begin{sidewaystable}
\caption{Evaluated models for predicting the risk of high-grade prostate cancer. Internal dataset of size n=678; validation dataset of size $\rm N_{test}$=1174; PCPThg, prostate cancer prevention trial risk calculator; ERSPC, the European Randomized Study of Screening for Prostate Cancer risk calculator 3; SE, the proposed bootstrap standard error from 500 replicates; grey represents certain predictor was not used in the external calculator, yellow represents population-specific effects that are different from the internal population, and red represents poor performance compared with direct regression.} 
\label{Table_realdata}
\small
{\setlength{\tabcolsep}{1pt}
\begin{tabular}{@{}llcccccccccccccccccc@{}}
\hline
\multirow{3}{*}{~}  
& \multirow{3}{*}{~} 
& \multicolumn{2}{l}{PCPThg} 
& \multicolumn{2}{l}{ERSPC} 
& Direct
& \multicolumn{6}{l}{Proposed method} 
\\
\cline{3-6} \cline{8-13} 
& ~ 
& \multirow{2}{*}{Original} 
& \multirow{2}{*}{Estimated} 
& \multirow{2}{*}{Original} 
& \multirow{2}{*}{Estimated} 
& regression
& \multicolumn{3}{l}{Different intercept only$\dagger$} 
& \multicolumn{3}{l}{Different intercept and covariates $\ddagger$}
\\
\cline{8-13}& ~ & ~ & ~ & ~ & ~ & (REF) & Internal & PCPThg  & ERSPC & Internal  & PCPThg & ERSPC
\\ 
\hline
~ 
&
Intercept
& -3.686
& -1.409 (.115)
& -3.16 
& -1.391 (.111)
& -4.202 (.455)
& -4.157 (.456)
& {\cellcolor[rgb]{1,0.949,0.8}}-6.884 (.516)
& {\cellcolor[rgb]{1,0.949,0.8}}-6.013 (.515)
& -4.178 (.483)
& {\cellcolor[rgb]{1,0.949,0.8}}-6.884 (.516)
& {\cellcolor[rgb]{1,0.949,0.8}}-5.953 (.504)
\\
&
$\rm log_2$(PSA)
& 0.894
& 0.735 (.124)
& 1.176
& 0.891 (.124)
& 0.885 (.146)
& {\cellcolor[rgb]{0.886,0.937,0.851}}1.082 (.070)
& 1.082 (.070)
& 1.082 (.070)
& 0.877 (.152)
& {\cellcolor[rgb]{1,0.949,0.8}}1.099 (.101)
& {\cellcolor[rgb]{1,0.949,0.8}}1.202 (.100)
\\
~ 
& DRE
& 1 
& 1.145 (.257)
& 1.813
& 1.306 (.269)
& 1.045 (.299)
& {\cellcolor[rgb]{0.886,0.937,0.851}}1.118 (.139)
& 1.118 (.139)
& 1.118 (.139)
& 1.044 (.310)
& {\cellcolor[rgb]{1,0.949,0.8}}0.707 (.229)
& {\cellcolor[rgb]{1,0.949,0.8}}1.614 (.184)
\\
Point
& Age
& 0.03         
& 0.033 (.012)
& {\cellcolor[rgb]{0.906,0.902,0.902}} 
& {\cellcolor[rgb]{0.906,0.902,0.902}} 
& 0.032 (.014)
& {\cellcolor[rgb]{0.886,0.937,0.851}}0.030 (.010)
& 0.030 (.010)
& {\cellcolor[rgb]{0.906,0.902,0.902}} 0.030 (.010)
& {\cellcolor[rgb]{0.886,0.937,0.851}}0.032 (.014)
& {\cellcolor[rgb]{1,0.949,0.8}}0.030 (.012)
& {\cellcolor[rgb]{0.906,0.902,0.902}}0.032 (.014)
\\
estimates
& Biopsy
& -0.36 
& -1.444 (.272)
& {\cellcolor[rgb]{0.906,0.902,0.902}}            
& {\cellcolor[rgb]{0.906,0.902,0.902}} 
& -1.201 (.294)
& {\cellcolor[rgb]{0.886,0.937,0.851}}-0.618 (.184)
& -0.618 (.184)
& {\cellcolor[rgb]{0.906,0.902,0.902}}-0.618 (.184)
& {\cellcolor[rgb]{0.886,0.937,0.851}}-1.195 (.303)
& {\cellcolor[rgb]{1,0.949,0.8}}-0.069 (.207)
& {\cellcolor[rgb]{0.906,0.902,0.902}}-1.195 (.303)
\\
(SE)
& Race
& 0.96  
& 0.442 (.288)
& {\cellcolor[rgb]{0.906,0.902,0.902}} 
& {\cellcolor[rgb]{0.906,0.902,0.902}}    
& 0.183 (.331)
& {\cellcolor[rgb]{0.886,0.937,0.851}}0.434 (.222)
& 0.434 (.222)
& {\cellcolor[rgb]{0.906,0.902,0.902}}0.434 (.222)
& {\cellcolor[rgb]{0.886,0.937,0.851}}0.186 (.360)
& {\cellcolor[rgb]{1,0.949,0.8}}0.680 (.258)
& {\cellcolor[rgb]{0.906,0.902,0.902}}0.186 (.360)
\\
~ 
& $\rm log_2$(TRUS-PV)
& {\cellcolor[rgb]{0.906,0.902,0.902}~}            
& {\cellcolor[rgb]{0.906,0.902,0.902}~} 
& -1.514
& -1.697 (.225)
& -1.699 (.254)
& {\cellcolor[rgb]{0.886,0.937,0.851}}-1.553 (.175)
& {\cellcolor[rgb]{0.906,0.902,0.902}}-1.553 (.175)
& -1.553 (.175)
& {\cellcolor[rgb]{0.886,0.937,0.851}}-1.656 (.271)
& {\cellcolor[rgb]{0.906,0.902,0.902}}-1.656 (.271)
& {\cellcolor[rgb]{1,0.949,0.8}}-1.443 (.611)
\\
& $\rm log_2$(PCA3+1)
& {\cellcolor[rgb]{0.906,0.902,0.902}}            
& {\cellcolor[rgb]{0.906,0.902,0.902}}            
& {\cellcolor[rgb]{0.906,0.902,0.902}}          
& {\cellcolor[rgb]{0.906,0.902,0.902}}        
& 0.495 (.089)
& 0.486 (.083)
& {\cellcolor[rgb]{0.906,0.902,0.902}}0.486 (.092)
& {\cellcolor[rgb]{0.906,0.902,0.902}}0.486 (.092)
& 0.494 (.094)
& {\cellcolor[rgb]{0.906,0.902,0.902}}0.494 (.094)
& {\cellcolor[rgb]{0.906,0.902,0.902}}0.494 (.094)
\\
& $\rm log_2$(T2:ERG+1)
& {\cellcolor[rgb]{0.906,0.902,0.902}}   
& {\cellcolor[rgb]{0.906,0.902,0.902}}   
& {\cellcolor[rgb]{0.906,0.902,0.902}}   
& {\cellcolor[rgb]{0.906,0.902,0.902}}
& 0.098 (.038)
& 0.092 (.037)
& {\cellcolor[rgb]{0.906,0.902,0.902}}0.092 (.037)
& {\cellcolor[rgb]{0.906,0.902,0.902}}0.092 (.037)
& 0.094 (.039)
& {\cellcolor[rgb]{0.906,0.902,0.902}}0.094 (.039)
& {\cellcolor[rgb]{0.906,0.902,0.902}}0.094 (.039)
\\ 
\hline
Prediction
& 
AUC
& 0.700
& 0.707
& 0.720
& 0.723
& 0.806 (REF)
&  0.804
&  0.804
&  0.804
&  0.805 
& {\cellcolor[rgb]{0.984,0.894,0.835}}0.797
&  0.802
\\
metrics
& Brier score 
& 1.059
& 0.987
& 0.954
& 0.994
& 0.903 (REF) 
& 0.915   
& {\cellcolor[rgb]{0.984,0.894,0.835}}0.977 
& {\cellcolor[rgb]{0.886,0.937,0.851}}0.875
& 0.902
& {\cellcolor[rgb]{0.984,0.894,0.835}}0.990
& {\cellcolor[rgb]{0.886,0.937,0.851}}0.875
\\
\hline
\multicolumn{20}{l}{
$\dagger$ \scriptsize{
    $\rm logit(p_i)= \rm \gamma_0 + \gamma_0^{\rm {PCPThg}} I_{PCPThg} + \gamma_0^{\rm {ERSPC}} I_{ERSPC} 
    + \rm \gamma_{X_1} \rm log_2(PSA_i) +\gamma_{X_2} DRE_i +\gamma_{X_3} Age_i + \gamma_{X_4} Biopsy_i +\gamma_{X_5} Race_i + \gamma_{X_6} log_2(\text{TRUS-PV}_i) + \gamma_{B_1} log_2(PCA3_i+1)+ \gamma_{B_2} log_2(\text{T2:ERG}_i+1)$ (*)
}}
\\
\multicolumn{20}{l}{
$\ddagger$ \scriptsize{
    $\rm logit(p_i)= (*)
    + I_{PCPThg}  \big[\gamma_1^{\rm {PCPThg}}log_2(PSA_i)+ \gamma_2^{\rm {PCPThg}}DRE_i+\gamma_3^{\rm {PCPThg}}Age_i+\gamma_4^{{PCPThg}} Biopsy_i+\gamma_5^{{PCPThg}} Race_i\big]
    + I_{ERSPC}  \big[\gamma_1^{{ERSPC}} \rm log_2(PSA_i)+\gamma_2^{{ERSPC}} DRE_i+\gamma_6^{{ERSPC}} log_2(\text{TRUS-PV}_i)\big]$
}}
\end{tabular}
}
\end{sidewaystable}

The grey empty blocks in Table \ref{Table_realdata} imply that these predictors were not used in that specific external model, and thus in the proposed method we assume they have the same coefficient as the internal population (grey blocks with values). Results in Table \ref{Table_realdata} show that (i) for the internal population, we gain precision of the estimated coefficients by incorporating external model information (smaller SE highlighted in green), e.g., in the different-intercept-only model, the bootstrap SE of $\rm log_2$(PSA) reduces from 0.146 to 0.070 while it reduces for DRE from 0.299 to 0.139, compared to direct regression using internal data only; for the external population, when allowing population-specific effects (yellow blocks), (ii) we do not expect to see large precision gain due to variance-bias trade-off, e.g., the precision gain of both $\rm log_2$(PSA) and DRE we see in the different-intercept-only model diminishes in the different-intercept-and-covariates model, and (iii) the point estimates of the external populations are different from the internal population.

In the prediction metrics row, we show AUC (higher value represents better discrimination) and scaled Brier score (smaller value means improved accuracy) calculated using the validation cohort, where the red blocks implies slightly worse overall predictive performance of PCPThg population compared to direct regression, e.g., 0.9\% reduction of AUC. This may be because that the validation cohort has a moderately different population than the training cohort as it has different baseline distribution as noted by \citet{Tomlins2015}, such as lower African Americans rate and higher DRE percentages, which may also explain why the fitted model for the European population has better performance on the validation data.

\section{Simulation Studies} \label{simulation}
While our proposed approach can handle any target model that belongs to the class of GLM, we focus on a binary outcome and logistic regression to evaluate the performance of SynDI and comparison methods. In all scenarios, the internal data is of size 200, while we vary the synthetic data size from one times the internal data size (i.e., $\rm r_1=r_2=1$ and N=$\rm n \times [1+\rm r_1+r_2$]=600) to ten times (i.e., $\rm r_1=r_2=10$ and N=$\rm n \times [1+\rm r_1+r_2$]=4200) for each external population. We compare SynDI with three methods, one benchmark method using the internal data only and two common approaches to analyze the combined dataset created in step 1 of Section \ref{framework} where both of them include outcome $\rm Y$ during imputation:
\begin{enumerate}
    \item Internal-data-only (direct regression): fit the target model using the internal data only without incorporating external information;
    \item Proposed method SynDI: we implement it through MICE in R software. For example, in Figure 1 step 1, the imputation models for $\rm X_2$, $\rm X_3$ and $\rm B$ are ($\rm X_2|X_1,X_3,B$), ($\rm X_3|X_1,X_2,B$), and ($\rm B|X_1,X_2,X_3$), respectively. Weights are calculated after imputation;
    \item Fully conditionally specification (FCS): imputation through FCS by specifying an imputation model for each missing predictor conditional on all the observed covariates and the outcome Y, and iteratively generate imputed values \citep{VanBuuren2006}. For example, in Figure 1 step 1, the imputation models for $\rm X_2$, $\rm X_3$ and $\rm B$ are ($\rm X_2|X_1,X_3,B,Y$), ($\rm X_3|X_1,X_2,B,Y$), and ($\rm B|X_1,X_2,X_3,Y$), respectively; 
    \item Imputation by ordered monotone blocks (IMB): a strategy that was designed to handle block-wise missingness \citep{Li2014}, the data structure we have in step 1. In our case, it sequentially imputes missing covariates starting with the one with the minimum missingness conditional on the observed data, outcome, and newly imputed data. We implement IMB through MICE by specifying a different imputation model compared with FCS. For example, in Figure 1 step 1, the imputation models for $\rm X_2$, $\rm X_3$ and $\rm B$ are ($\rm X_2|X_1,Y$), ($\rm X_3|X_1,X_2,Y$), and ($\rm B|X_1,X_2,X_3,Y$), respectively.
\end{enumerate}

M=100 imputations are used for all multiple imputation. For FCS and IMB, we fit the same target model as SynDI but without weights, and calculate the variance via Rubin's combining rules \citep{Rubin2002}. 

\subsection{Simulation Settings}
We provide two representative examples in Simulation I and II to illustrate how to handle the two categories of external summary-level information, respectively (see summarized simulation settings in Figure \ref{sim_settings}). Additional simulation results to assess various settings and violations of assumptions can be found in Web Supplemental Section 4.

\begin{sidewaysfigure}
\centering
\adjustimage{width=.8\linewidth, center=1\linewidth}{simulation_settings}
\caption{Simulation settings snapshot.} \label{sim_settings}
\end{sidewaysfigure}

\begin{itemize}
    \item \textbf{Simulation I:} Idealized case where the internal data contains $\rm Y, X_1, X_2, B_1 \text{ (continuous)}$, $\rm B_2 \text{ (binary)}$, and two external models have been fitted to very large datasets that is sampled from the true data generating mechanism. The external models provided parameter estimates $\rm \hat{\boldsymbol{\beta}}_1$ and $\rm \hat{\boldsymbol{\beta}}_2$ from logistic regression models $\rm Y|X_1$ and $\rm Y|X_1, X_2$, respectively. In all three populations, $\rm X_1, X_2$ and $\rm B_1$ follows a standard multivariate normal distribution with zero-mean, standard deviation 1, and 0.3 for all pairwise correlations, while $\rm B_2$ follows a Bernoulli distribution $\rm B_2|X_1,X_2,B_1 \sim Ber([1+exp^{-1}(0.1X_1+0.2X_2+0.3B_1)])$. As shown in Figure \ref{sim_settings}, the three populations have similar generative outcome models with different intercepts, i.e., -1, 1 and 3, which give prevalence of Y=1 of 0.3, 0.57, and 0.81, respectively. The target model is a logistic regression with different intercepts of the form $\rm logit[Pr(Y=1|\mathbf{X, B, I})]= \gamma_0 + \gamma_0^{1}I_1 + \gamma_0^{2}I_2 + \gamma_{X_1} X_1+\gamma_{X_2}X_2+\gamma_{B_1}B_1+\gamma_{B_2}B_2$. 
    
    \it{Evaluation metrics:} \rm Absolute bias, the estimated variance from proposed bootstrap method and other comparisons, and the empirical variance of point estimates.
    
    \item \textbf{Simulation II:} Compared to simulation I, the external model 2 in simulation II is derived by fitting a random forest model to a large dataset where the underlying true generative model is a logistic regression model that contained quadratic and interaction terms. Specifically, the internal study contains complete data of $\rm Y, X_1, X_2, X_3, X_4$, $\rm B_1 \text{ (continuous)}$, $\rm B_2 \text{ (binary)}$, and the two external models are available in different forms of summary information. External model 1 provides $\rm \boldsymbol{\hat{\beta}}_1$ from a logistic regression model $\rm Y|X_1, X_2, X_3$ and external model 2 provides the estimated probabilities of Y=1 given $\rm X_1, X_2, X_3$ and $\rm X_4$ through a fitted random forest model. In all three populations, $\rm X_1, X_2$ and $\rm X_3$ follows a standard multivariate normal distribution with zero-mean, standard deviation 1. For all pairwise correlations, while $\rm X_4$ and $\rm B_1$ each follows a conditional normal distribution, $\rm X_4|X_1,X_2,X_3 \sim N(0.2\sum_{p=1}^3X_p, 1)$ and $\rm B_1|\mathbf{X} \sim N(0.2\sum_{p=1}^3X_p+0.1 X_4, 1)$, and $\rm B_2$ follows a Bernoulli distribution $\rm B_2|\mathbf{X},B_1 \sim Ber\big(\{1+exp^{-1}[0.2\sum_{p=1}^3X_p+0.1(X_4+B_1)]\}\big)$, respectively. Similar to simulation I, the true generative distributions of Y in the internal and external population 1 shared the same main covariate effect but have different intercepts (-1 and 2, which corresponds to prevalence 0.3 and 0.65, respectively), while external model 2 additionally contains a quadratic term and an interaction (with intercept 3 corresponding to prevalence 0.73). The target model is a logistic regression with the form $\rm logit[Pr(Y=1|\mathbf{X, B, I})]= \gamma_0 + \gamma_0^{1}I_1 + \gamma_0^{2}I_2 + \sum_{p=1}^4\gamma_{X_p} X_p+\sum_{q=1}^2\gamma_{B_q}B_q$. As described in step 3 of Section \ref{framework}, $\rm \hat{\boldsymbol{\beta}}_1$ can be directly used to calculate weights for $\rm I_1$ while we need to estimate $\rm \boldsymbol{\beta}^{syn}_2$ to calculate weights for $\rm I_2$. To obtain $\rm \hat{\boldsymbol{\beta}}^{syn}_2$, we first generate a large synthetic data set $\rm (\hat{Y}_2^{syn}, \mathbf{X}_2^{syn})$ by replicating the observed ($\rm X_1, X_2, X_3, X_4$) and generating $\rm \hat{Y}_2^{syn}$ values through the available random forest model, and then fit a main effect logistic model $\rm \hat{Y}_2^{syn}|\mathbf{X}_2^{syn}$ using only the synthetic data and ignoring the missing $\rm B_1$ and $\rm B_2$. 
    
    \it{Evaluation metrics:} \rm Three prediction metrics over a validation data of size $\rm N_{test}=2000$: Sum of squared error (SSE) $\rm = \frac{1}{N_{test}}\sum_{i=1}^{N_{test}}(\hat{p}_i-p_{i0})^2$, where $\rm \hat{p}_i$ and $\rm p_{i0}$ denotes the estimated and true probability of $\rm Y_i=1$ given $\rm \mathbf{X}_i$ and $\rm \mathbf{B}_i$, respectively; AUC and BS. 
\end{itemize}

\subsection{Simulation Results}
\begin{figure}[H]
\centering
\begin{subfigure}[t]{0.83\textwidth}
\centering
\includegraphics[width=\textwidth]{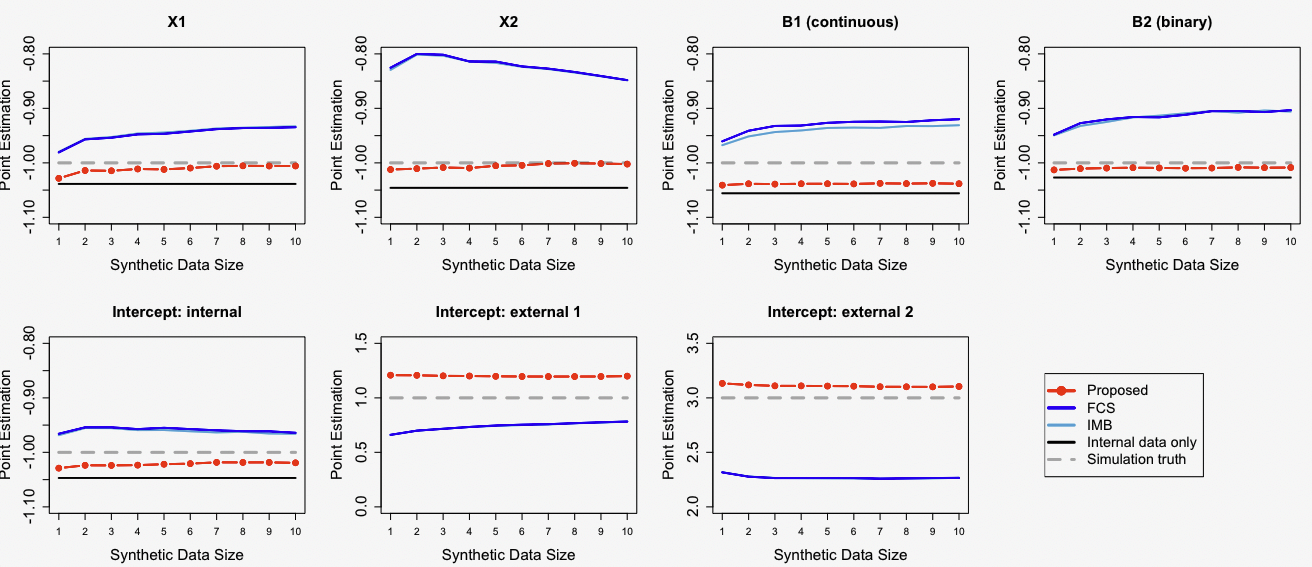} 
\caption{Point estimates of the target parameter $\rm \boldsymbol{\gamma}$} \label{sim1_A}
\end{subfigure}

\begin{subfigure}[t]{0.83\textwidth}
\centering
\includegraphics[width=\textwidth]{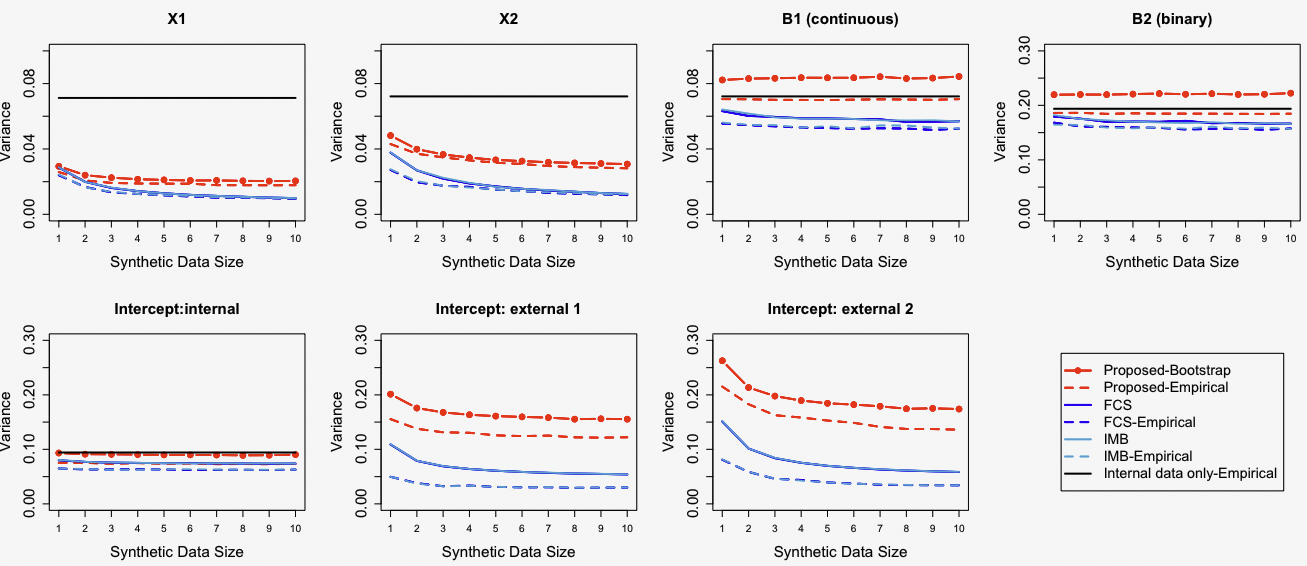} 
\caption{Variance estimators vs. the empirical variance of $\rm \hat{\boldsymbol{\gamma}}$} \label{sim1_B}
\end{subfigure}

\begin{subfigure}[t]{0.83\textwidth}
\centering
\includegraphics[width=\textwidth]{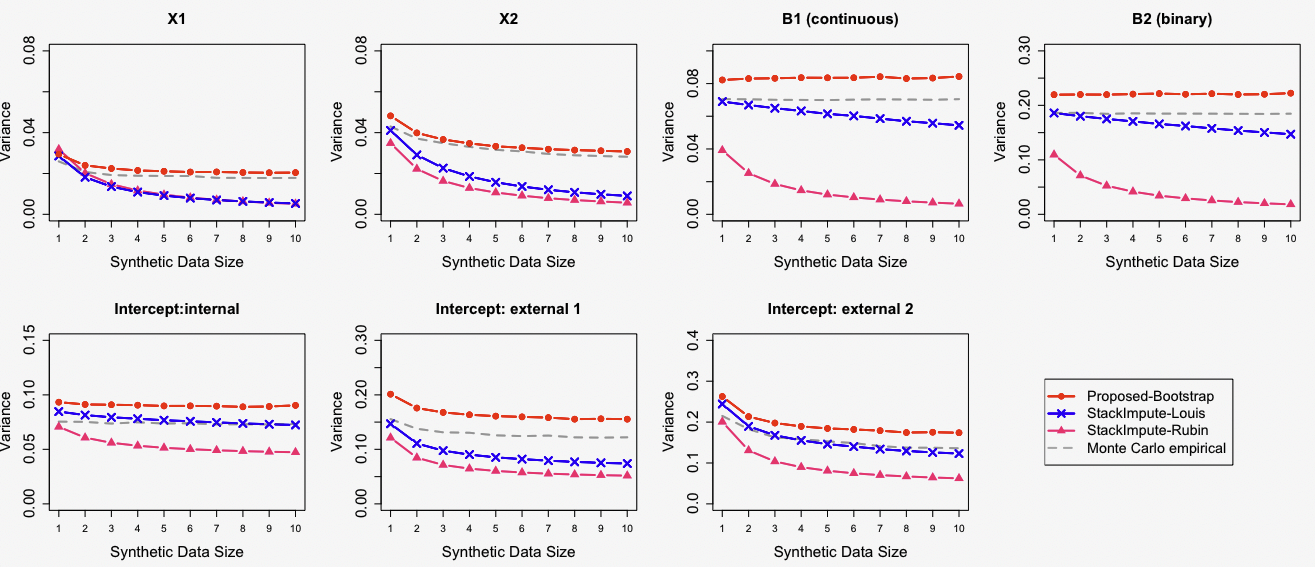} 
\caption{Comparison of the proposed and the existing variance estimators of $\rm \hat{\boldsymbol{\gamma}}$} \label{sim1_C}
\end{subfigure}

\caption{Visualization of simulation I results over increasing synthetic data size (a) point estimates of $\rm \boldsymbol{\gamma}$ (b) variance estimators vs. the empirical variance of $\rm \hat{\boldsymbol{\gamma}}$ (c) comparison of the proposed and the existing variance estimators of $\rm \hat{\boldsymbol{\gamma}}$.} \label{sim1}
\end{figure}

Figure \ref{sim1} shows the average results of estimating $\rm \boldsymbol{\gamma}$ from the target model across 500 simulated datasets for simulation I, including point estimates $\rm \hat{\boldsymbol{\gamma}}$ in Figure \ref{sim1_A}, variance estimators versus the empirical variance of $\rm \hat{\boldsymbol{\gamma}}$ in Figure \ref{sim1_B}, and the comparison of different variance estimators of $\rm \hat{\boldsymbol{\gamma}}$ in Figure \ref{sim1_C}. This figure appears in color in the electronic version of this article, and color refers to that version. Figure \ref{sim1_A} shows that FCS (dark blue curve) and IMB (light blue curve) have similarly biased estimates, indicating these traditional imputation strategies involving the outcome variable cannot distinguish heterogeneous population effects, while SynDI (red dotted curve) always shows close results to the simulation truth (grey dashed curve) for all covariates, especially estimating $\rm X_2$ and intercept of external population 2 when other methods show severe bias. 
Compared to the benchmark method (internal-data-only estimates in black solid curve), SynDI provides more precise point estimates and additional estimation for the external populations.

In Figure \ref{sim1_B}, each color denotes one method with a pair of curves, one solid curve representing the variance estimator and another dashed curve representing the Monte Carlo empirical variance of the point estimates. If the variance is correctly estimated, the solid curve should be approximately equal to the corresponding dashed one, which is true for all methods. All methods show precision gain in estimated X coefficients compared to the internal-data-only (the longer the distance to the internal-data-only curve, the larger the precision gain) while no precision gain in B covariates and the intercept due to no external added information and allowing population-specific effects, respectively. SynDI has over 50\% efficiency gain in estimated X coefficients compared to the internal data. We see FCS and IMB have larger precision gain in both estimated X coefficients than the proposed curve, which may be explained by bias-variance trade-off as they also have larger bias in the corresponding point estimates in Figure \ref{sim1_A}. We will discuss its statistical rationale in Section \ref{Discussion}.

Figure \ref{sim1_C} shows the result of the proposed variance estimator and two comparisons, the Rubin's rule variance estimator (StackImpute-Rubin) and the Louis information estimator (StackImpute-Louis), the variance estimators proposed by \cite{beesley2021stacked}. \cite{Gu2019} has shown that when the synthetic data size goes to infinity, the precision gain we achieve in X covariates will converge, which is shown as the gradually stable trend of the grey curve (Monte Carlo empirical variance of the point estimates, also serves as the empirical truth here). When the synthetic data size increases, 
the StackImpute-Louis estimator continuously underestimate the empirical truth of the variance estimates for all coefficients, and the StackImpute-Rubin estimator has similar pattern in estimating the variance of coefficients for $\rm X, B$ and the intercept of external study 1. On the contrary, the proposed variance estimator is empirically accurate, especially in X covariates where the empirical bias of variance in other methods can be up to ten times higher than the proposed method (e.g., 0.02 versus 0.002 in absolute bias), the bias of which could be even larger when the synthetic data size keep increasing. Moreover, the proposed variance estimator is robust with increasing synthetic sample size whereas it increases for the other variance estimators. Note that the internal-data-only results (black solid curve) does not exist in external intercepts as they were not available in the internal data, whereas the bias of the internal data estimates is due to the small sample bias compared with the simulation truth.

\begin{figure}[H]
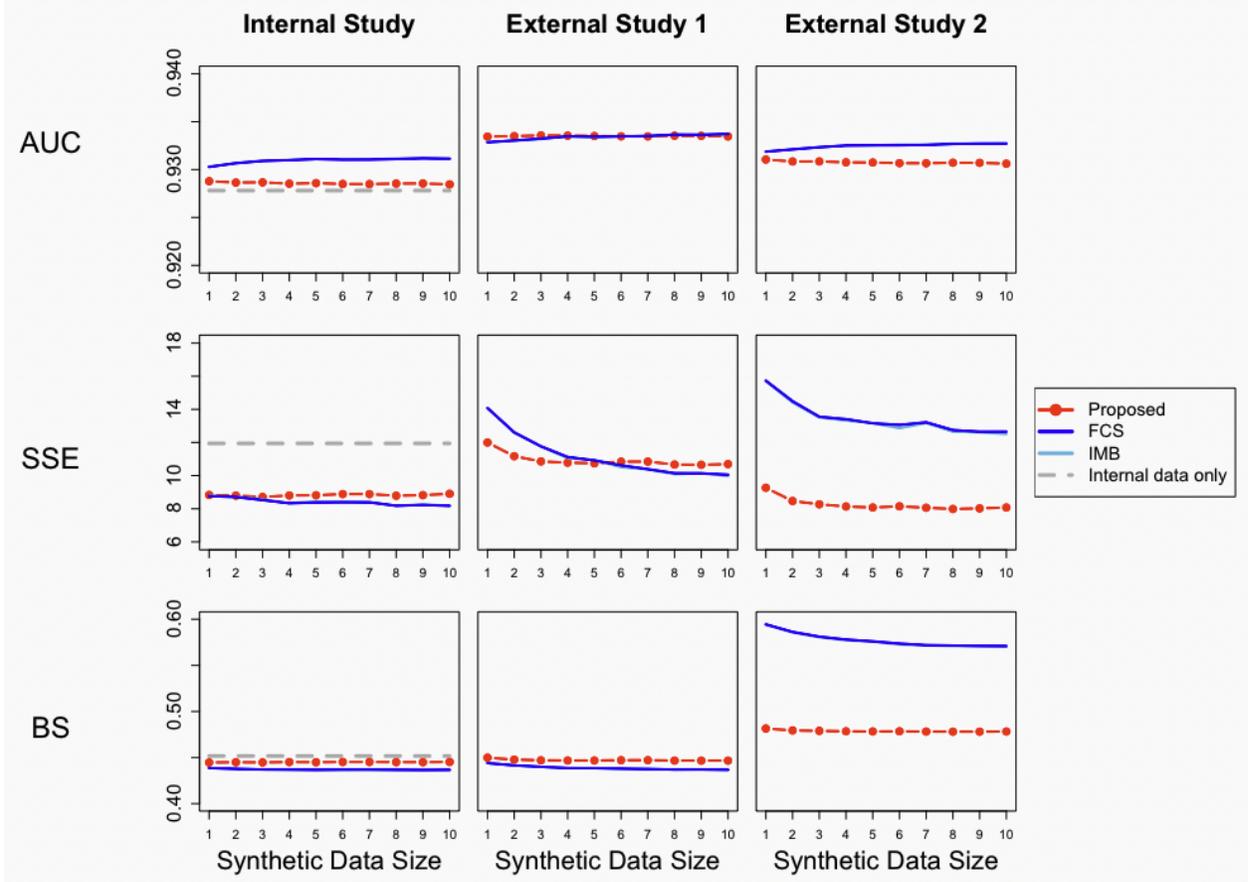

\centering
\adjustimage{width=1\linewidth, center=1\linewidth}{sim2}
\caption{Visualization of prediction metrics over increasing synthetic data size for simulation II. Larger AUC (area under the curve), smaller SSE (sum of squared error) and smaller BS (Brier Score) represents better prediction accuracy.} \label{sim2}
\end{figure}

Figure \ref{sim2} shows the result of Simulation II over increasing synthetic data size on a validation dataset, with three prediction metrics in the rows and three populations in the columns. In general, the results in Figure \ref{sim2} are in line with Figure \ref{sim1_A}, implying that SynDI has better overall prediction performance compared with others. Specifically, the first column indicates that all methods incorporating external information have consistently better prediction ability (larger AUC, smaller SSE and smaller BS) than the internal-data-only method (the dashed grey line). While all methods have similar performance in terms of AUC and predicting internal population, SynDI outperforms others in terms of SSE and BS in predicting external populations, especially external study 2 where the true parameter values are quite different from the internal study values (SynDI has up to 41\% more improvement in SSE and 19\% more improvement in BS compared with FCS and IMB). SynDI shows a modest improvement in performance as the size of the synthetic data increases, e.g., for $\rm I_2$, SSE decreases 12.9\% from 9.3 to 8.1 in SynDI when the synthetic data size increases from one to ten. Note that it is hard to distinguish FCS and IMB in the figure as they have very close results.

\section{Discussion} \label{Discussion}
\textbf{Flexibility in external models and populations:} The proposed approach adds to the existing research on integrating external summary information into the internal study. It can develop improved models and provide statistical inference not only for the internal population but also for the external populations. In the target model, the parameters that were measured in the external reduced models are allowed to differ from those of the internal population. This new strategy has the appealing feature of utilizing external information that comes in the form of a ``black box" algorithm, i.e., an algorithm that provides a predicted probability, but the underlying model is not necessarily simple or transparent or even known. The key aspect is that the external information allows the creation of synthetic data. We further summarize some key points and concluding remarks of the proposed strategy in the subsequent paragraphs.

\textbf{Using partial information through data integration:} The proposed method can integrate summary information from multiple external models that each uses different covariates $\rm \mathbf{X}_k \in \mathbf{X}$ into the current study. The simulation and real data analysis showed expected results that we can only gain precision on estimated coefficient of $\rm \mathbf{X}$ but not the $\rm \mathbf{B}$ coefficients that are only available internally, even when $\mathbf{X}$ and $\mathbf{B}$ are correlated. This is consistent with the theoretical results in \cite{Dai2012} that the MLE estimator $\rm \hat{\boldsymbol\beta}$ and $\rm \hat{\boldsymbol\gamma}_B$ are always asymptotically independent under regularity conditions, where $\rm \hat{\boldsymbol\beta}$ is the estimates of intercept and $\rm \mathbf{X}$ coefficients in model $\rm Y|\mathbf{X}; \boldsymbol\beta$ and $\rm \hat{\boldsymbol\gamma}_B$ is the estimated coefficient of $\rm \mathbf{B}$ from model $\rm Y|\mathbf{X, B}; \boldsymbol\gamma$, respectively. 

\textbf{Principled inference post-imputation:} Although we discuss the framework under GLM, the proposed strategy borrows strengths from the stacked multiple imputation (StackImpute) proposed by \citet{beesley2021stacked}, which can avoid incompatibility between the imputation model and the analysis model, and can be easily extended to accommodate complicated outcomes such as the time-to-event outcome in survival models. In the proposed method, we introduce a bootstrap-typed variance estimation and compare it with the analytical variance estimators proposed in \citet{beesley2021stacked, beesley2021accounting}. In general, the bootstrap variance can provide more valid variance estimation but may be more computationally intense than others, while analytical estimates are fast to compute but may be biased. Based on simulation results, when the predictors have small covariate effects (simulation in Section 3.2 of Web Supplemental), the bias of the analytical variance estimates is small. 

\textbf{Improve not just current study but external model predictions:} To our knowledge, very few existing approaches can allow different population effects through regression analysis in data fusion, let alone improving external model predictions. Beyond regression analysis, several approaches were proposed in the causal inference and survival analysis field, aiming for a similar goal to incorporate the supplementary information from the validation dataset to the main dataset and allowing heterogeneous treatment effects among different data sources \citep{Dominici2017, Wang2012, Yang2020, Huang&Qin2018, Chen2020}. The proposed framework also allows readers to decide which covariate effects to differ across populations. Such decisions, allowing only intercept to differ or allowing all possible covariate effects to differ, depend on readers' prior knowledge about cross-population effects and research emphasis on variance-bias trade-off (i.e., more flexible model will lead to potentially less bias and larger variance estimation).

\textbf{Allow for violation of transportability assumption:} The transportability assumption is common in data integration and causal inference when certain predictors are not mutually observed across populations \citep{Rassler2004, Reiter2012, Bareinboim2013}. Compared with more strict transportability assumed in literature, e.g., transportability of the joint $\rm Y,\mathbf{X, B}$ \citep{Chatterjee2016} and conditional transportability of $\rm Y| \mathbf{X, B}$ or $\rm \mathbf{B}| \mathbf{X}, Y$ \citep{Dominici2017, Estes2017, Gu2021}, we only require conditional transportability among covariates $\rm \mathbf{X}$ and $\rm \mathbf{B}$ (Assumption 2). While the simulation results (Section 3.4 of Web Supplemental) suggest that violating this assumption could have mild impact, one can consider applying additional shrinkage methods such as the empirical Bayes approaches proposed by \citet{Estes2017} and \citet{Gu2021} after obtaining estimates from the proposed approach, which can empirically strike a balance between bias and efficiency when the transportability between populations is unclear.

\textbf{Limitations and future directions:} 
The proposed method assumes that the covariates that were not used in the external studies have the same effect as the internal study. 
Therefore, caution must be exercised when interpreting external populations' results as the true underlying relationship is hard to verify without additional data from external populations. If possible, it is recommended to cross-validate the results using additional data sources from external populations. An interesting extension of the proposed method is to accommodate the situation where selection bias exists and selection probability or survey weights are available for each observation in the internal population. In theory, the proposed method can be adapted to accommodate this by replacing the synthetic Y values with the inverse probability-weighted or survey weights-weighted synthetic Y values in step 1 of the proposed strategy. Alternatively, instead of copying the whole internal X's multiple times to create the same X distribution as the internal population, one can consider proportionally creating synthetic X through the given weights to recover the representative distribution in the external populations. Further investigation is needed to evaluate this. Furthermore, if the exposure indicator is available as a covariate in all populations, one can also use the regression estimates from the proposed method to calculate the estimated average causal effect by averaging over the joint distribution of ($\rm \mathbf{X, B}$).

\backmatter

\section*{Acknowledgements}
This research was supported by the NIH grants CA129102, R01HG008773-05, 1UG3CA267907-01 and 5P30CA046592-30; and NSF grant DMS1712933. The authors thank Dr. Lauren Beesley for many helpful suggestions that led to a significant improvement of this paper. \vspace*{-8pt}

\section*{Data Availability Statement}
Data from the illustrative example are not shared due to third-party data sharing restrictions and to protect patient privacy.

\section*{Supporting Information}
The Web Supplemental Sections referenced in Section \ref{motivation}, \ref{framework}, \ref{simulation} and \ref{Discussion} are available online. An R package SynDI implementing the proposed method can be found on GitHub at\\ https://github.com/umich-biostatistics/SynDI, which will soon be available on CRAN.

\bibliographystyle{biom}
\bibliography{refs.bib}

\label{lastpage}

\end{document}


\maketitle

\linespread{2}

\section{Baseline Characteristics of High-Grade Prostate Cancer Datasets}

\begin{table}[H]
\centering
\begin{adjustbox}{width=1\textwidth}
\begin{tabular}{llllll}
\hline
 &  & \textbf{Internal data} & \textbf{Validation data} & \textbf{PCPThg}  & \textbf{ERSPC}       \\
\hline
\multicolumn{2}{l}{Sample size}  & 678     & 1174            & 5519        & 3616        \\
\multicolumn{2}{l}{Number of case (prevalence)} & 179 (26.4\%)  & 214 (18.2\%)    & 257 (4.7\%) & 313 (35.4)  \\
\hline
Median (range) &  PSA (ng/mL) & 5.1 (0.3-460.4)    & 4.6 (0.1-290.0)      & 1.5 (0.3–287.0)           & 4.3 (0.1–316.0)     \\
\hline
\multirow{3}{*}{Mean (SD)} 
& Age (years) & 62.6 (8.4)    & 64.0 (8.9)      & 69.7 (Unknown)    & 65.5 (Unknown)    \\
& PCA3  & 38.2 (40.3)   & 45.2 (65.7)     & Unknown           & Unknown           \\
&  T2:ERG  & 77.7 (325.9)  & 72.6 (365.0)    & Unknown           & Unknown           \\
\hline
\multirow{3}{*}{Count (\%)} & Abnormal DRE & 96 (14.2)     & 283 (24.1)      & Unknown           & 1280 (35.4) \\
& Prior biopsy  & 167 (24.6)    & 235 (20.0)      & 753 (13.6)  & Unknown   \\
& African American  & 70 (10.3)     & 68 (5.8)        & 175 (3.2)   & Unknown  \\
\hline                                  
\end{tabular}
\end{adjustbox}
\end{table}

\section{Assumptions of the proposed approach}
To implement the multiple imputation in step 2, some covariate information need to be shared across populations, since the missing covariates are completely unobserved in one population, also known as block-wise missing structure. Specifically, assumption 2 in Figure 3 in the main text states that the conditional distribution of the missing covariates to be the same across populations given the observed covariates (i.e., $\rm \mathbf{X}_{miss} \perp\!\!\!\perp [I_0,I_1,...,I_K] | \mathbf{X}_{obs}$; and $\rm \mathbf{B} \perp\!\!\!\perp [I_0,I_1,...,I_K]|\mathbf{X}$). Therefore, the imputation models are $\rm f(X_{miss}|\mathbf{X}_{obs})$ and $\rm f(B|\mathbf{X})$ for missing $\rm \mathbf{X}$ and missing $\rm \mathbf{B}$, respectively (e.g., $\rm \mathbf{X}_{miss}=[X_2, X_3]$ and $\rm X_{obs}=X_1$ in Figure 1 in the main text). 

\section{Deriving the Initial Estimates for the External Populations}
In this section, we will show how to obtain the initial parameter estimates of external population k. Let $\rm (\hat{\gamma}_0, \boldsymbol{\rm \hat{\gamma}}_X^{T}, \boldsymbol{\rm \hat{\gamma}}_B^{T})^T$ be the internal data only estimates of $\rm Y|\mathbf{X, B}, I_0$ using internal data only. For external population k, we know the parameter estimates $\rm \hat{\boldsymbol{\beta}}_k=(\hat{\beta}_0, \boldsymbol{\rm \hat{\beta}}_{X_k}^T)^T$ from the fitted model $\rm Y|\mathbf{X}_k; \boldsymbol{\beta}_k$. We assume that all predictors, $\rm \mathbf{X}$ and $\rm \mathbf{B}$, are centered, and the true target model parameter for the external population k is $\rm (\gamma_0^{k}, \boldsymbol{\rm \gamma}_X^{kT}, \boldsymbol{\rm \gamma}_B^{T})^T$, assuming the coefficient of the unobserved variable $\rm \mathbf{B}$ is the same as the internal population. 

The goal of estimating $\rm \gamma_0^{k}$ and $\rm \boldsymbol{\rm \gamma}_X^{k}$ from model $\rm Y|\mathbf{X, B}, I_k; \boldsymbol{\gamma}_{I_k}$ is equivalent to correcting the bias of $\rm \hat{\boldsymbol{\beta}}_k$ in the reduced model $\rm Y|\mathbf{X}_k; \boldsymbol{\beta}_k$ considering covariates $\rm \mathbf{X}_{(-k)}$ and $\rm \mathbf{B}$ as omitted. To simplify notation, we assume $\rm \mathbf{B}$ is the only omitted covariate 
in the derivation below. Neuhaus and Jewell (1993) provided a Taylor-series-expansion approximation to show that the ratio of coefficients remains constant in both the reduced and the full model when the omitted $\rm \mathbf{B}$ is independent of the observed $\rm \mathbf{X}$, i.e. $\rm \frac{\gamma_{X_1}}{\gamma_{X_2}} \approx \frac{\beta_{X_1}}{\beta_{X_2}}$, indicating that the relative effect size among regression coefficients remains consistent across models. In their Table 3 and equation 9 [1] provided the algebraic relationship between $\rm \boldsymbol{\rm \gamma}_X^{k}$ and $\rm \boldsymbol{\rm \beta}_X$ for exponential family when the omitted $\mathbf{B}$ and the observed $\mathbf{X}$ are correlated. In the subsequent paragraphs, we will explain in detail how to estimate $\rm \gamma_0^{k}$ and $\rm \boldsymbol{\rm \gamma}_X^{k}$ in linear regression (continuous Y) and logistic regression (binary Y), respectively.

\medskip
\textbf{1. Linear Regression:} Suppose E($\rm \mathbf{B} | \rm \mathbf{X}; \boldsymbol\theta$)= $\rm \boldsymbol{\rm \theta X}$. We start by replacing $\rm \mathbf{B}$ with the conditional expected value E($\rm \mathbf{B} | \rm \mathbf{X}; \boldsymbol\theta$) in the mean profile of the target model:
\begin{align*}
      \rm E(Y|\mathbf{X, B}; \boldsymbol\gamma) &= \rm \gamma_0^{k} + \boldsymbol{\rm \gamma}_X^{k} \mathbf{X} + \boldsymbol{\rm \gamma}_B\mathbf{B}
      = \rm \gamma_0^{k} + \boldsymbol{\rm \gamma}_X^{k} \mathbf{X} + \boldsymbol{\rm \gamma}_B\boldsymbol{\theta} \mathbf{X}
    = \rm E(Y|\mathbf{X}; \boldsymbol{\gamma}, \boldsymbol{\theta}) 
\end{align*}
Since $\rm \hat{E}(Y|\mathbf{X}; \boldsymbol{\beta}) = \hat{\beta}_0 + \boldsymbol{\rm \hat{\beta}}_X \mathbf{X}$ is available through the externally fitted model, we can obtain the estimation of $\rm \gamma_0^{k}$ and $\rm \boldsymbol{\rm \gamma}_X^{k}$ by matching the intercept and $\rm \mathbf{X}$ coefficient between $\rm \hat{E}(Y|\mathbf{X};\boldsymbol{\gamma}, \boldsymbol{\theta})$ and $\rm \hat{E}(Y|\mathbf{X}; \boldsymbol{\beta})$, respectively: $\rm \hat{\gamma}_0^{k} = \hat{\beta}_0$ and $\rm \boldsymbol{\rm \hat{\gamma}}_X^{k} = \boldsymbol{\rm \hat{\beta}}_{X_k} - \boldsymbol{\rm \theta}^T\boldsymbol{\rm \hat{\gamma}}_B $. In a special case where the internal and the external population only differ in intercept, we can directly obtain the initial estimates $\rm \hat{\boldsymbol{\rm \gamma}}_{I_k}=(\rm \hat{\beta}_0, \hat{\boldsymbol{\rm \gamma}}_X^{T}, \hat{\boldsymbol{\rm \gamma}}_B^{T})^T$.

\medskip
\textbf{2. Logistic Regression:}
In logistic regression where g() is the logit link function, we connect the intercepts $\rm \beta_0$ and $\rm \gamma_0^{k}$ through the equation $\rm logit^{-1}(\beta_0) = E_{B|X}(\mu_0^{k})$, where $\rm \mu_{0}^{k}=g^{-1}(Y|\mathbf{X}, \mathbf{B}; \boldsymbol{\rm\gamma}_X^{k}=0)=logit^{-1}(\gamma_0^{k}+\mathbf{B}^T\boldsymbol{\rm \gamma}_B)$. For the right hand side, we expand $\rm \mathbf{B}$, a vector of length Q, at $\rm E(\mathbf{B|X})$ using the third-order Taylor series expansion as follows:

\begin{align*}
        \rm E_{B|X}(\mu_0^{k}) &= \rm E_{B|X}[logit^{-1}(\gamma_0^{k}+\mathbf{B}^T\boldsymbol{\rm \gamma}_B)] 
        \\
        &\approx \rm logit^{-1}(w) \Big\{1+\frac{1}{2}\frac{1-e^{w}}{(1+e^{w})^2}\sum_{i=1}^Q\sum_{j=1}^Q\gamma_{B_i}\gamma_{B_j} E_{B|X}\Big[\Big(B_i-E(B_i|\mathbf{X})\Big)\Big(B_j-E(B_j|\mathbf{X})\Big)\Big]\Big\}
        \\
        &= \rm logit^{-1}(w) \Big[1+\frac{1}{2}\frac{1-e^w}{(1+e^w)^2}Var\Big(\sum_{i=1}^Q\gamma_{B_i}B_i|\mathbf{X}\Big)\Big] \numberthis \label{taylor_expansion_logit-1}
\end{align*}
where $\rm w = \gamma_0^{k}+E(\mathbf{B}^T|\mathbf{X})\boldsymbol{\rm \gamma}_B$. Given $\rm \hat{\beta}_0, \boldsymbol{\rm \hat{\gamma}}_B, \hat{E}(\mathbf{B|X})$ and $\rm \hat{V}ar(\mathbf{B|X})$, we can easily obtain $\rm \hat{\gamma}_0^{k}$ by solving the equation $\rm E_{B|X}(\mu_0^{k}) - logit^{-1}(\hat{\beta}_0)=0$.

After obtaining $\rm \boldsymbol\gamma_0^{k}$, we then estimate $\rm \boldsymbol\gamma_X^{k}=(\gamma_{X_1}^{k},...,\gamma_{X_{P_k}}^{k})^T$ according to the following equation provided in Neuhaus and Jewell (1993):

\setlength{\abovedisplayskip}{0pt} \setlength{\abovedisplayshortskip}{0pt}
\begin{equation*}
    \rm \boldsymbol{\beta}_{X_p} = \rm \Big\{\boldsymbol{\gamma}_{X_p}^{k}+\Big[E(\mathbf{B}^T|\mathbf{X}+1_p)-E(\mathbf{B}^T|\mathbf{X})\Big]\boldsymbol{\gamma}_B \Big\}\Big\{1-\frac{Var_{B|X}(\mu_0^{k})}{1-E_{B|X}(\mu_0^{k})[1-E_{B|X}(\mu_0^{k})]}\Big\}
\end{equation*}
where $\rm 1_p$ is a zero vector with the $\rm p^{th}$ term equals to 1 and $\rm p \in \{1,...,P_k\}$. Similar to equation (\ref{taylor_expansion_logit-1}), we can also obtain the Taylor-series-expansion estimation for $\rm E_{B|X}[(\mu_0^{k})^2]=E_{B|X}[logit^{-2}(\gamma_0^{k}+\boldsymbol{\rm B}^T\boldsymbol{\rm \gamma}_B)] \approx \rm \frac{e^{2w}}{(1+e^w)2} \Big[1+\frac{1}{2}\frac{2-e^w}{(1+e^2)^2}\sum_{i=1}^Q\sum_{j=1}^Q \gamma_{B_i} \gamma_{B_j} Cov(B_i,B_j|\mathbf{X})\Big]$, together with $\rm E_{B|X}(\mu_0^{k})$, we then obtain an approximation of $\rm V_{B|X}(\mu_0^{k})=E_{B|X}[(\mu_0^{k})^2]-E_{B|X}(\mu_0^{k})$. Given $\rm \boldsymbol{\rm \hat{\beta}}_X$, $\rm \boldsymbol{\rm \hat{\gamma}}_B$, $\rm \hat{E}(\mathbf{B|X})$, and $\rm \hat{\gamma}_0^{k}$, we can obtain $\rm \boldsymbol{\rm \hat{\gamma}}_X^{k}=\boldsymbol{\rm \hat{\beta}}_X(1-\frac{\hat{V}_{B|X}(\mu_0^{k})}{\hat{E}_{B|X}(\mu_0^{k})[1-\hat{E}_{B|X}(\mu_0^{k})]})^{-1}-\Big[\hat{E}(\mathbf{B}^T|\mathbf{X}+1_p)-\hat{E}(\mathbf{B}^T|\mathbf{X})\Big]\hat{\boldsymbol{\gamma}}_B$.

Note that we estimate $\rm E(\mathbf{B|X};\boldsymbol{\theta})=\rm g'^{-1}(\boldsymbol\theta\mathbf{X})$ and $\rm Var(\mathbf{B|X};\boldsymbol{\theta})=g'^{-1}(\boldsymbol\theta\mathbf{X})\Big[1-g'^{-1}(\boldsymbol\theta\mathbf{X})\Big]$ using the internal data by regressing each B on $\rm \mathbf{X}$ with appropriate link function $\rm g'()$ based on the type of B, e.g., when B is continuous, linear regression and identity link is used; when B is binary, logistic regression and logit link is used. Given $\rm \hat{\boldsymbol{\theta}}$, $\rm \hat{E}(\mathbf{B|X})=\hat{\boldsymbol{\theta}}^T E(\mathbf{X})$ is used.

\section{Additional Simulation Results}
In this section, we show the results of additional simulations to assess the performance of the proposed strategy for point estimates and variance estimation.

\subsection{Continuous outcome Y (a supplement to Simulation I in the main manuscript)}
\textbf{Goal:} To examine the proposed method when the outcome is continuous and the target model is linear regression.
\\
\noindent \textbf{Simulation setup:} This simulation is the same as Simulation I in the main manuscript, except the generative outcome model now follows Gaussian distribution:
\begin{align*}
        \begin{cases}
         \text{Internal: } &\rm Y|\mathbf{X, B} \sim N(-1 - X_1 - X_2 - B_1 - B_2, 1);\\
         \text{External 1: } &\rm Y|\mathbf{X, B} \sim N(1 - X_1 - X_2 - B_1 - B_2, 1);\\
         \text{External 2: } &\rm Y|\mathbf{X, B} \sim N(3 - X_1 - X_2 - B_1 - B_2, 1).
        \end{cases}
    \end{align*}
The target outcome model (model 2 in the main manuscript) is now a linear regression:
\begin{equation*}
    \rm E(Y|\mathbf{X, B, S}) = \rm \gamma_{0} + \sum_{k=1}^2 \gamma_0^{S_k}S_k + \sum_{p=1}^2 \gamma_{X_p} X_p + \sum_{q=1}^2 \gamma_{B_q} B_q,
\end{equation*}

\begin{figure} 
    \centering
    \begin{subfigure}[t]{0.77\textwidth}
        \centering
        \includegraphics[width=\textwidth]{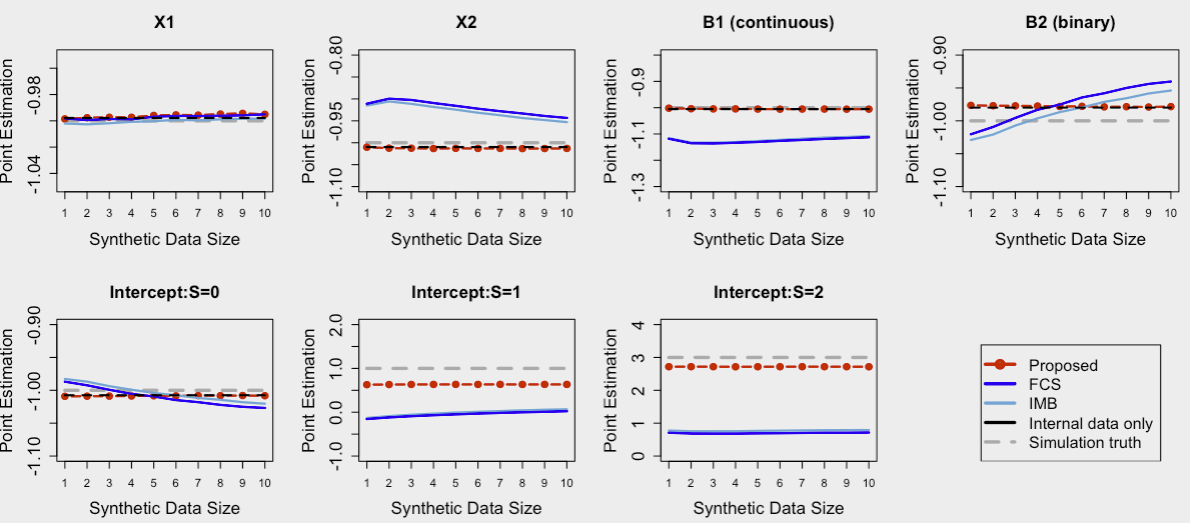} 
        \caption{Point estimates} 
        \label{simS1_A}
    \end{subfigure}
    
    \begin{subfigure}[t]{0.77\textwidth}
        \centering
        \includegraphics[width=\textwidth]{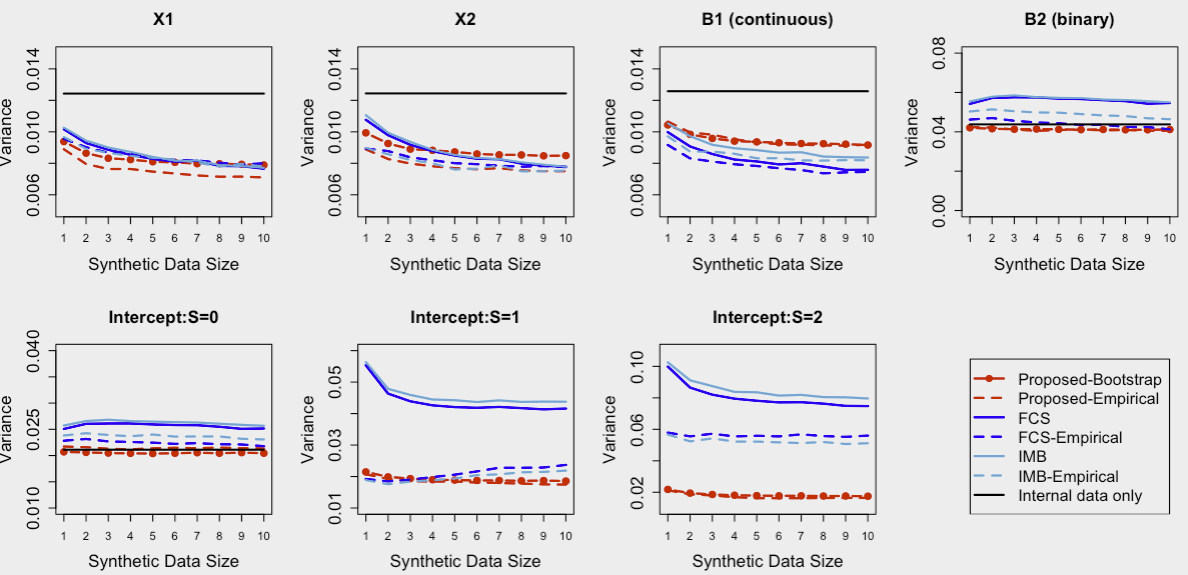} 
        \caption{Variance estimator vs. the empirical variance} 
        \label{simS1_B}
    \end{subfigure}
    
    \begin{subfigure}[t]{0.77\textwidth}
        \centering
        \includegraphics[width=\textwidth]{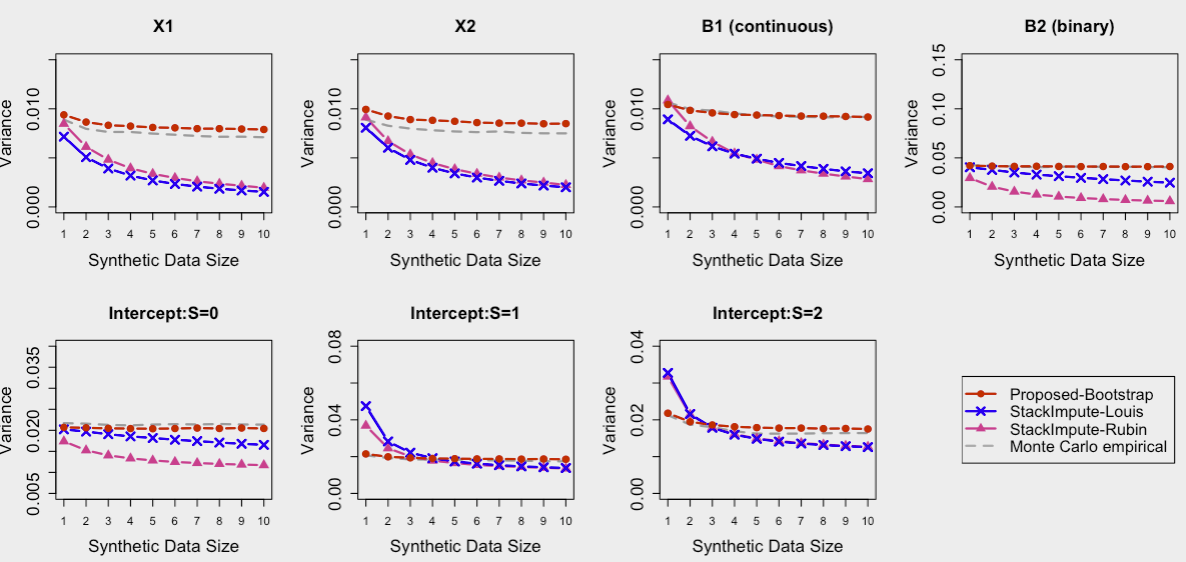} 
        \caption{Different variance estimators of the proposed method} 
        \label{simS1_C}
    \end{subfigure}
    
    \caption{Results of Simulation 1.1 over increasing synthetic data size (a) point estimates (b)variance estimation vs. Monte Carlo empirical variance (c) different variance estimators of the proposed method.} \label{simS1}
\end{figure}

\noindent \textbf{Results:} Figure \ref{simS1} shows similar pattern as those in Simulation I in the main manuscript, where the proposed method (red dotted curve) has the smallest bias among all for all covariates (Figure \ref{simS1_A}), largest precision gain compared with others (Figure \ref{simS1_B}), and the closest variance estimation to the Monte Carlo empirical variance (Figure \ref{simS1_C}).

\subsection{Smaller covariate effect (a modification to Simulation I in the main manuscript)}
\begin{figure}[H]
    \centering
    \begin{subfigure}[t]{0.8\textwidth}
        \centering
        \includegraphics[width=\textwidth]{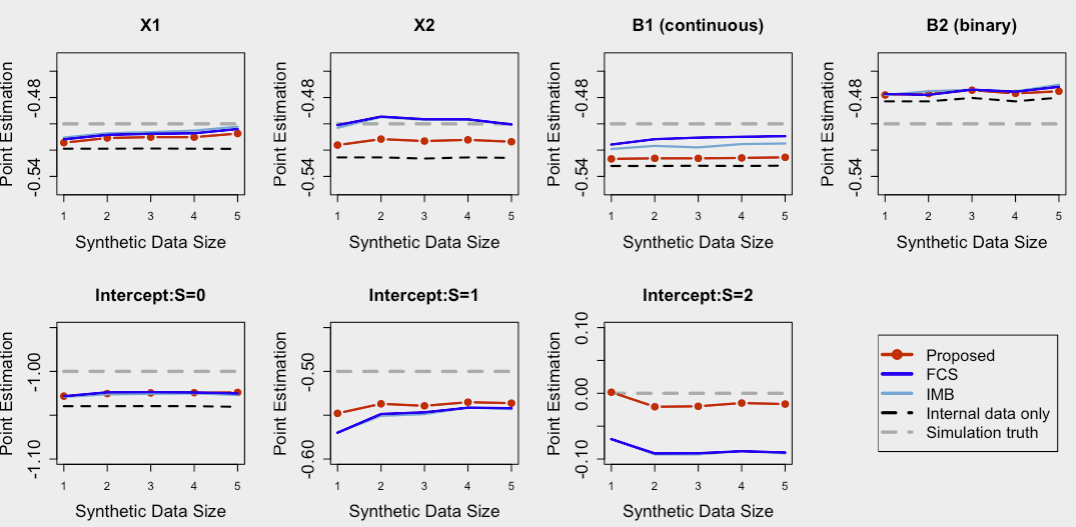} 
        \caption{Point estimates} 
        \label{simS2_A}
    \end{subfigure}
    
    \begin{subfigure}[t]{0.8\textwidth}
        \centering
        \includegraphics[width=\textwidth]{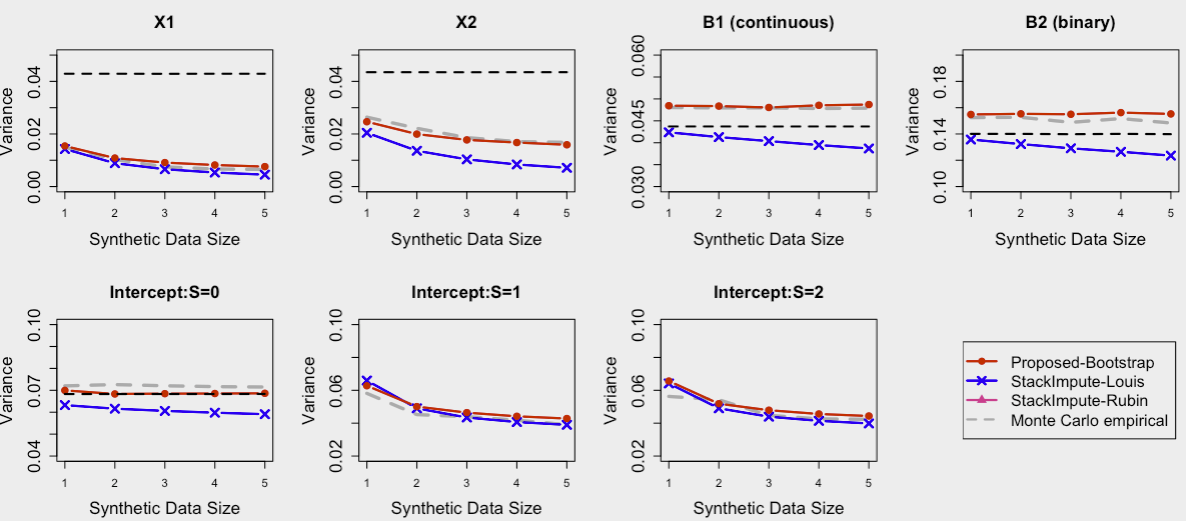} 
        \caption{Different variance estimators of the proposed method}
        \label{simS2_B}
    \end{subfigure}
    
    \caption{Results of Simulation 1.2 over increasing synthetic data size (a) point estimates (b) different variance estimators of the proposed method.} \label{simS2}
\end{figure}

\textbf{Goal:} To assess our approach when the magnitude and the difference of covariate effects are small across different populations in the target outcome model.
\\
\noindent \textbf{Simulation setup:} This simulation is the same as Simulation I in the main manuscript, except the coefficient effect is now -0.5 instead of -1, and the intercept difference is smaller among populations:
\begin{align*}
        \begin{cases}
         \text{Internal: } &\rm logit[Pr(Y=1|\mathbf{X, B})] = -1 - 0.5\big(X_1 + X_2 + B_1 + B_2\big),  \text{ prevalence=0.28};\\
         \text{External 1: } &\rm logit[Pr(Y=1|\mathbf{X, B})] = -0.5 - 0.5\big(X_1 + X_2 + B_1 + B_2\big),  \text{ prevalence=0.36};\\
         \text{External 2: } &\rm logit[Pr(Y=1|\mathbf{X, B})] = 0 - 0.5\big(X_1 + X_2 + B_1 + B_2\big),  \text{ prevalence=0.45}.
        \end{cases}
    \end{align*}
\\
\noindent \textbf{Results:} Figure \ref{simS2_A} shows that compared with larger covariate effects in Simulation I in the main manuscript, when the X covariate effect is small, FCS and IMB have smaller bias in estimating X coefficients but still lack the ability to identify population-specific effects (i.e. intercepts of external populations). Similarly, Figure \ref{simS2_B} shows smaller bias of variance estimation. Note that the Rubin's rule variance estimator in Figure \ref{simS2_B} is too large (the pink curve) so that it falls outside of the range of the figure.

\subsection{Different X covariate effects in the outcome model (a more flexible outcome model compared with Simulation I in the main manuscript)}
\textbf{Goal:} In the main manuscript, we only present the simulation results allowing the target model's intercept to differ across populations. In this simulation, we additionally show the performance of the proposed method when all possible X covariates coefficients are allowed to differ across populations (similar to model 1 or ``different intercept and covariates" model in the real data example in the main manuscript). 

\noindent \textbf{Simulation setup:} This simulation is the same as Simulation I in the main manuscript except now that the generative outcome models are as follows:
\begin{align*}
        \begin{cases}
         \text{Internal: } &\rm logit[Pr(Y=1|\mathbf{X, B})] = -1 - X_1 - X_2 - B_1 - B_2, \text{ prevelance= 0.3};\\
         \text{External 1: } &\rm logit[Pr(Y=1|\mathbf{X, B})] = 1 + X_1 - X_2 - B_1 - B_2, \text{ prevelance= 0.58};\\
         \text{External 2: } &\rm logit[Pr(Y=1|\mathbf{X, B})] = 3 + 3X_1 + 3X_2 - B_1 - B_2,
          \text{ prevelance= 0.70}.
        \end{cases}
    \end{align*}

\begin{figure}[H]
     \centering
     \begin{subfigure}[b]{0.49\textwidth}
         \centering
         \includegraphics[width=\textwidth]{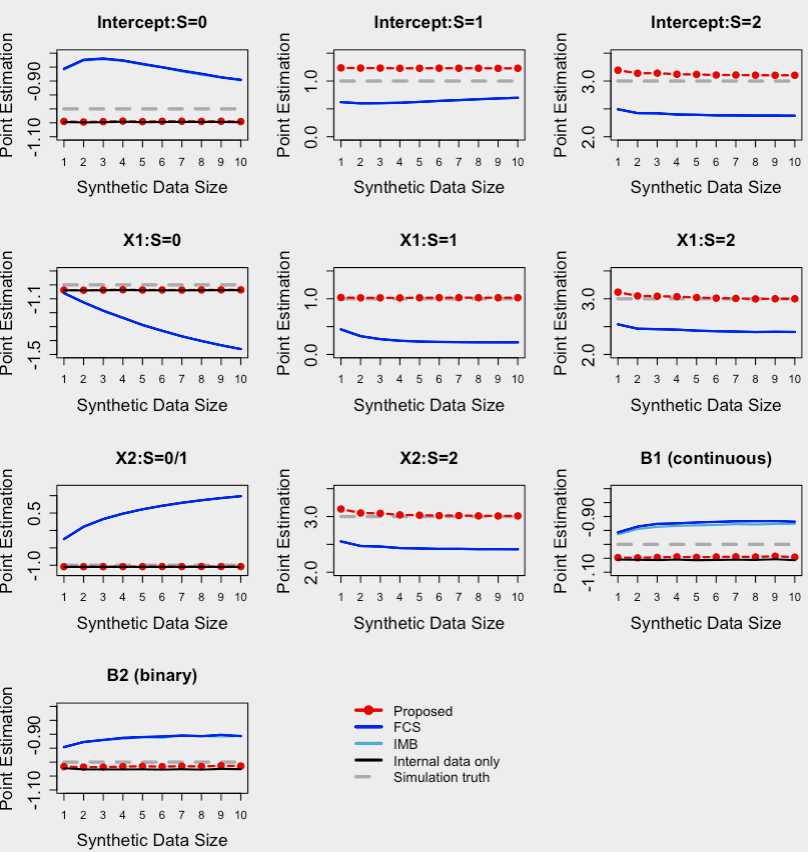}
         \caption{Point estimates}
         \label{simS3_A}
     \end{subfigure}
     \hfill
     \begin{subfigure}[b]{0.49\textwidth}
         \centering
         \includegraphics[width=\textwidth]{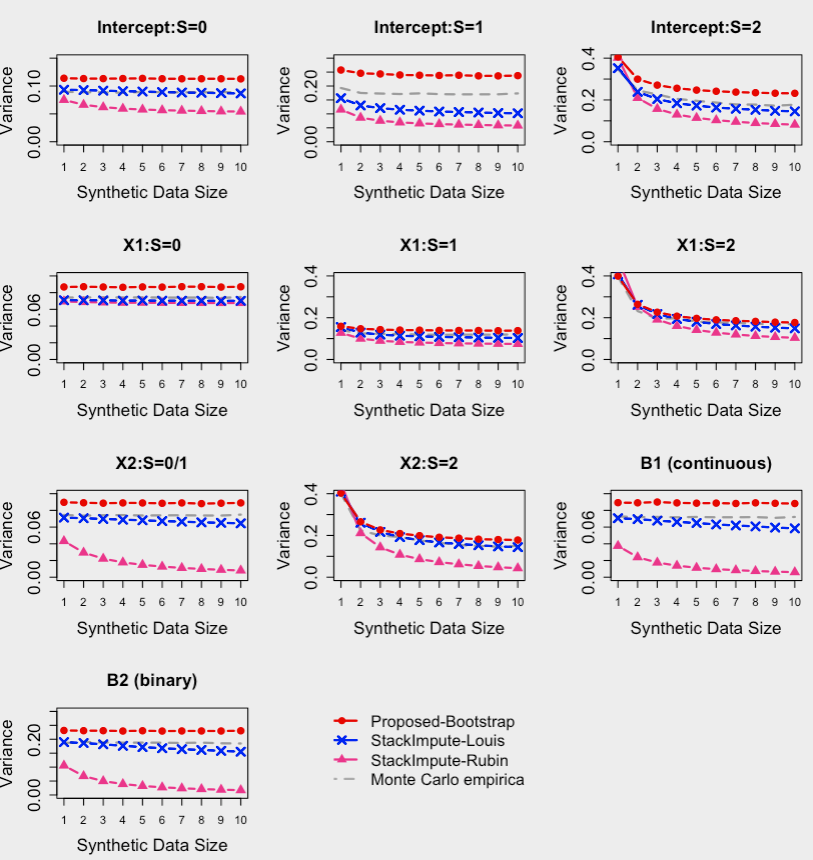}
         \caption{Different variance estimators of the proposed method}
         \label{simS3_B}
     \end{subfigure}
        \caption{Results of Simulation 1.3 over increasing synthetic data size (a) point estimates (b) different variance estimators of the proposed method.}
        \label{simS3}
\end{figure}

\noindent \textbf{Results:} Similar to the results of Simulation I in the main manuscript, the results in Figure \ref{simS3} shows outstanding performance of the proposed method in both point estimates and variance estimation compared with others. For example, the proposed method has small bias less than 0.02 when estimating $\rm X_2 $ in population S=1 while the bias in FCS and IMB can go up to 0.78 (i.e. almost 40 times of the proposed method).

\subsection{Violation of transportability assumption}
\textbf{Goal:} To examine the proposed method when Assumption 2 ($\rm X_{miss}|X_{obs}$ and $\rm B|X$ are transportable between the internal and the external populations) is violated. We present two examples where the violation only causes ignorable bias in case 1 while it has larger impact in case 2. 

\subsubsection{\textbf{Case 1: different $\rm B|X$ distribution in external population 2}} \textbf{Simulation setup:} This simulation is the same as Simulation I in the main manuscript except that now the external model 2 has different marginal $\rm B_1$ distribution and different conditional distribution $\rm B_2|X_1, X_2, B_1$:
\begin{itemize}
    \item $\rm B_1$ has mean 1.5 and standard deviation 1.5 in external population 2 while in other populations $\rm B_1$ has mean 0 and standard deviation 1;
    \item $\rm B_2|\mathbf{X},B_1 \sim Ber\{[1+exp^{-1}(0.2X_1+0.3X_2+0.4B_1)]\}$ in external population 2 while in other populations $\rm B_2|\mathbf{X},B_1 \sim Ber\{[1+exp^{-1}(0.1X_1+0.2X_2+0.3B_1)]\}$.
\end{itemize}
Note that both $\rm B_1$ and $\rm B_2$ are only observed in the internal study and multiple imputations are needed for them, where $\rm B_2|\mathbf{X}$ and $\rm B_2|\mathbf{X},B_1$ should be the same across populations according to Assumption 2.

\begin{figure}[H]
     \centering
     \begin{subfigure}[b]{0.6\textwidth}
         \centering
         \includegraphics[width=\textwidth]{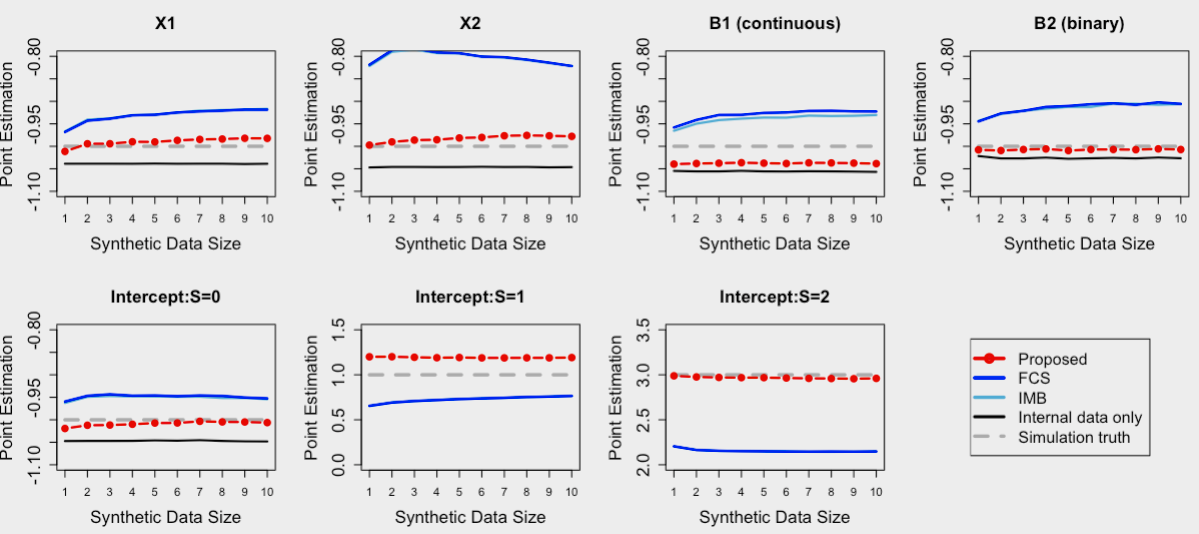}
         \caption{Point estimates}
         \label{simS4.1_A}
     \end{subfigure}
     \hfill
     \begin{subfigure}[b]{0.6\textwidth}
         \centering
         \includegraphics[width=\textwidth]{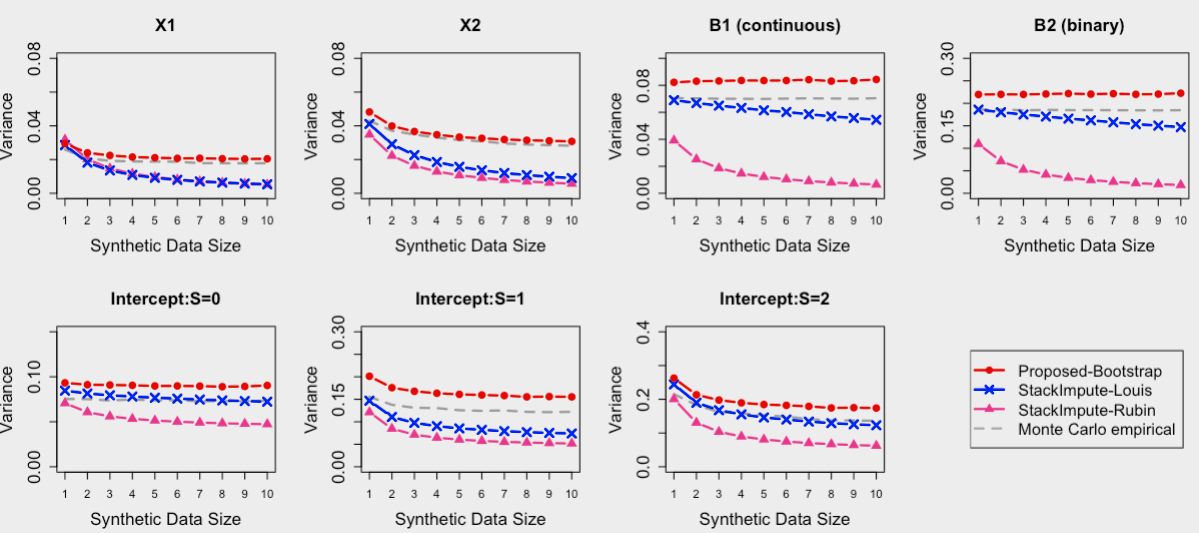}
         \caption{Different variance estimators of the proposed method}
         \label{simS4.1_B}
     \end{subfigure}
        \caption{Results of Simulation 1.4.1 over increasing synthetic data size (a) point estimates (b) different variance estimators of the proposed method.}
        \label{simS4.1}
\end{figure}

\noindent \textbf{Results:} Figure \ref{simS4.1_A} indicates that the violation of transportability assumption in the proposed method has limited impact of point estimation with ignorable bias while Figure \ref{simS4.1_B} shows similar pattern of variance estimations as before. 

\subsubsection{\textbf{Case 2: different marginal $\rm X_1$ distribution in external populations}} \textbf{Simulation setup:} This simulation is the same as Simulation I in the main manuscript except now that in the external studies, $\rm X_1 \sim N(1, 1.5)$ while in the internal study $\rm X_1 \sim N(0, 1)$. This will lead to different distribution conditional on $\rm X_1$ and thus violates Assumption 2.

\begin{figure}[H]
    \centering
    \begin{subfigure}[b]{0.8\textwidth}
        \centering
        \includegraphics[width=\textwidth]{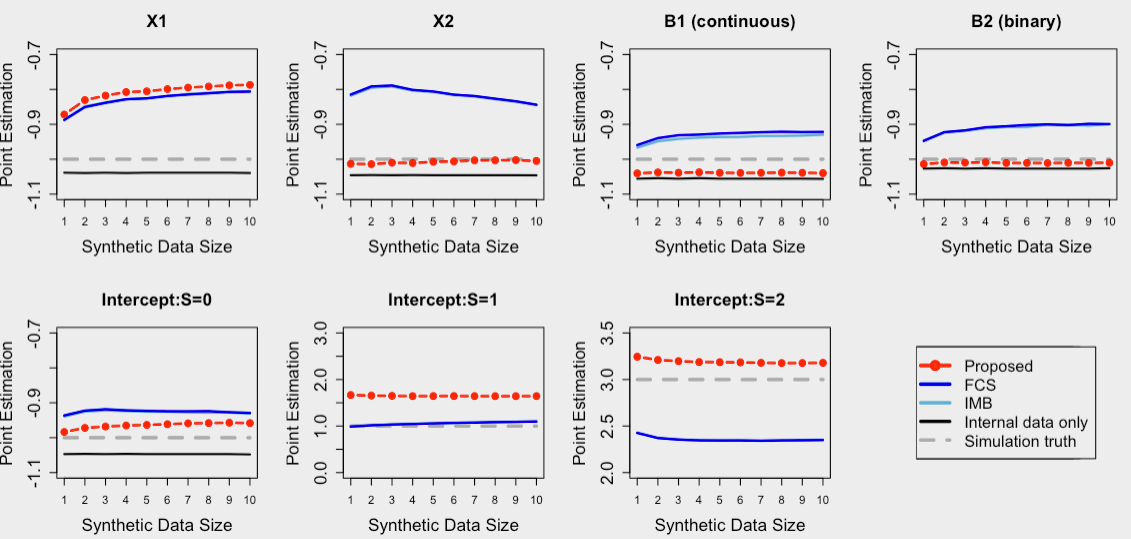} 
        \caption{Point estimates} 
        \label{simS4.2_A}
    \end{subfigure}
    
    \begin{subfigure}[b]{0.8\textwidth}
        \centering
        \includegraphics[width=\textwidth]{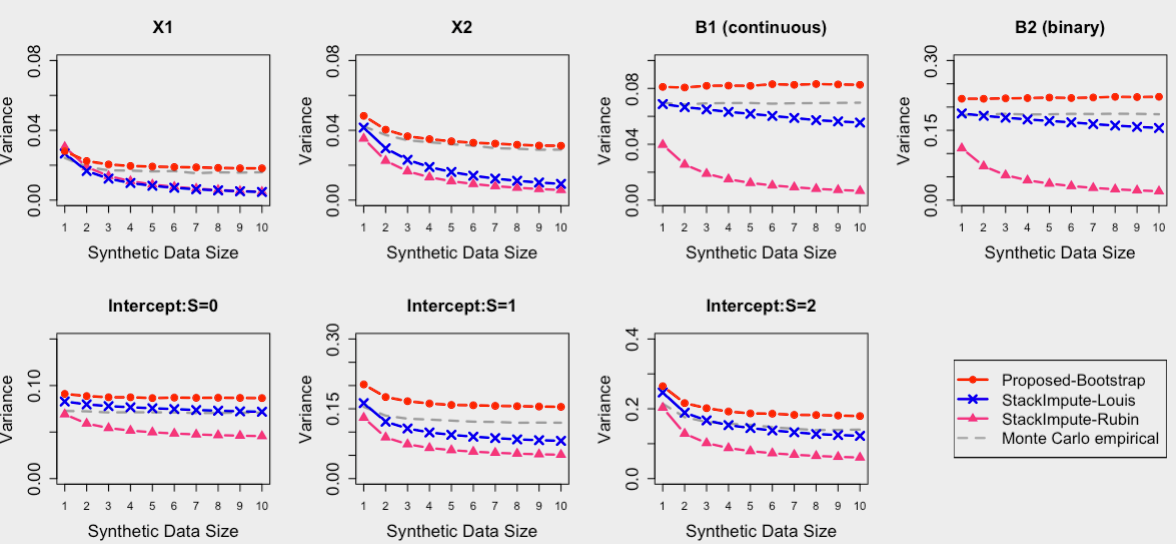} 
        \caption{Different variance estimators of the proposed method}
        \label{simS4.2_B}
    \end{subfigure}
    
    \caption{Results of Simulation 1.4.2 over increasing synthetic data size (a) point estimates (b) different variance estimators of the proposed method.} \label{simS4.2}
\end{figure}

\noindent \textbf{Results:} Figure \ref{simS4.2_A} shows that such violation leads to some bias of estimated coefficient $\rm X_1$, i.e., 0.2 absolute bias. Besides that, the proposed method has nearly unbiased point estimates for other parameter (i.e. up to 0.014 absolute bias) while the bias in FCS and IMB can be up to 15 times the bias of the proposed method. Similarly, Figure \ref{simS4.2_B} shown unbiased variance estimation of the proposed bootstrap estimator.

